\documentclass[final,aps,prd,twocolumn,showpacs,showkeys, 10pt, longbibliography, nolinenumbers, floatfix]{revtex4-2}

\usepackage{amsmath}
\usepackage{amssymb}
\usepackage{amsfonts}
\usepackage{graphicx}
\usepackage{float}
\usepackage{dcolumn}
\usepackage{multirow}
\usepackage{natbib}
\usepackage{tcolorbox}
\usepackage[colorlinks = true,linkcolor = blue,urlcolor  = blue,citecolor = blue,anchorcolor = blue]{hyperref}
\usepackage{enumerate}
\usepackage{siunitx}
\usepackage{stfloats} 
\usepackage[expansion=true, protrusion=true]{microtype}
\usepackage{orcidlink}
\usepackage{times}
\usepackage{siunitx}

\begin{document}

\title{Configurational entropy and stability conditions of fermion and boson stars }

\author{P.S. Koliogiannis$^{1,2}$\orcidlink{0000-0001-9326-7481}}
\email{pkoliogi@auth.gr}
\author{M. Vikiaris$^2$\orcidlink{0009-0005-8779-4556}}
\email{mvikiari@auth.gr}
\author{C. Panos$^2$\orcidlink{0000-0002-4604-0433}}
\email{chpanos@math.auth.gr}
\author{V. Petousis$^3$\orcidlink{0000-0002-5575-6476}}
\email{Vlasios.Petousis@cvut.cz}
\author{M. Veselsk\'y$^3$\orcidlink{0000-0002-7803-0109}}
\email{Martin.Veselsky@cvut.cz}
\author{Ch.C. Moustakidis$^2$\orcidlink{0000-0003-3380-5131}}
\email{moustaki@auth.gr}
\affiliation{$^1$Department of Physics, Faculty of Science, University of Zagreb, Bijeni\v cka cesta 32, 10000 Zagreb, Croatia \\
$^2$Department of Theoretical Physics, Aristotle University of Thessaloniki, 54124 Thessaloniki, Greece\\
$^3$Institute of Experimental and Applied Physics, Czech Technical University, Prague, 110 00, Czechia}

\begin{abstract}
In a remarkable study by M. Gleiser and N. Jiang [Phys. Rev. D {\bf 92}, 044046, 2015], the authors demonstrated that the stability regions of neutron stars, within the framework of the simple Fermi gas model, and self-gravitating configurations of complex scalar field (boson stars) with various self couplings, obtained through traditional perturbation methods, correlate with critical points of the configurational entropy with an accuracy of a few percent. Recently, P. Koliogiannis \textit{et al.} [Phys. Rev. D {\bf 107}, 044069 2023] found that while the minimization of the configurational entropy generally anticipates qualitatively the stability point for neutron stars and quark stars, this approach lacks universal validity. In this work, we aim to further elucidate this issue by seeking to reconcile these seemingly contradictory findings.  Specifically, we calculate the configurational entropy of bosonic and fermionic systems, described by interacting Fermi and boson gases, respectively, that form compact objects stabilized by gravity. We investigate whether the minimization of configurational entropy coincides with the stability point of the corresponding compact objects. Our results indicate a strong correlation between the stability points predicted by configurational entropy and those obtained through traditional methods, with the accuracy of this correlation showing a slight dependence on the interaction strength. Consequently, the stability of compact objects, composed of components obeying Fermi or boson statistics, can alternatively be assessed using the concept of configurational entropy.


\end{abstract}

\maketitle
\section{Introduction}
In recent years there has been an extensive interest in the study of astrophysical objects with the help of the concept of information entropy and related quantities. In particular,  Sañudo and Pacheco~\cite{Sanudo-2009} studied the  relation between the complexity and the structure of white dwarfs. Later on,  the aforementioned study has been applied in neutron star's structure~\cite{Moustakidis-2009} and  it was found that the interplay between gravity, the short-range nuclear force, and the very short-range weak interaction shows that neutron stars, under the current theoretical framework, are ordered systems. Similar studies took place in the following years in a series of papers~\cite{de_Avellar-2012,de_Avellar-2014,Adhitya_2020,Contreras-2021,Posada-2021} and Herrera \textit{et al.}~\cite{Herrera-2018a,Herrera-2018b,Herrera-2019a,Herrera-2019b} elaborated the definition of the  complexity factors in self-gravitating systems, approaching the problem in a different way. Some additional applications of the concept of the  information measures may be found in 
Refs.~\cite{Sharif-2018a,Sharif-2018b,Sharif-2019,Sharif-2022,Yousaf-2020a,Yousaf-2020b,Yousaf-2020c,Yousaf-2021,Yousaf-2022}.    

A more specific application of the information measure is the configurational entropy (CE). The concept of the CE has been introduced by Gleiser and Stamatopoulos~\cite{Gleiser-2012} in order to study possible relation between the dynamical and information content of physics models with localized energy configurations. In the next years, the CE has been applied in several similar studies~\cite{Gleiser-2013,Gleiser-2015a,Koliogiannis-2023,Gleiser-2015b,Braga-2019,Braga-2020,Yunes-2018,Rocha-2021,Karapetyan-2018,Correa-2016,Baretto-2022}. In a notable study, Gleiser and Jiang~\cite{Gleiser-2015a} investigated the connection between  the stability of compact objects (white dwarfs, neutron stars and boson stars) and the corresponding information-entropic measure. According to their notable finding, the minimization of the CE offers an alternative way to predict the stability condition through the maximum mass configuration, for a variety of stellar objects. It is worth noting that there is no theoretical argument (or proof) to relate the stability point to the minimum of the CE. However, it is intuitive to expect that since the maximum mass corresponds to the most compact configuration (maximum mass and minimum radius of a stable configuration), the corresponding CE will exhibit an extreme value (in this case a total minimum).

Recently in Ref.~\cite{Koliogiannis-2023} we extended the aforementioned study in various compact objects including  neutron stars, quark stars, and hybrid stars. Employing a large set of realistic equations of state (EoS), in each case, we found that  the suggested prediction of the stability by the minimization of the CE, concerning neutron stars and quark stars, does not have, at least quantitatively, universal validity. 

It is worth mentioning that the study of the  longstanding problem of the stability of relativistic stars~\cite{Shapiro-1983,Glendenning-2000,Haensel-2007,Zeldovich-71,Weinberg-72,Schutz-85,Bielich-2020}  is mainly carried out by the following three methods: (a) the method of locating the point that corresponds to the minimization of the binding energy defined as $E_B=(M-m_bN)c^2$ (where $m_b$ is the mass of a single nucleon, $M$ stands for the gravitational mass, and $N$ is the total number of nucleons)~\cite{Shapiro-1983,Glendenning-2000,Haensel-2007,Zeldovich-71,Weinberg-72,Schutz-85,Bielich-2020}, (b) the variational method developed by Chandrasekhar~\cite{Chandrasekhar-1964a,Chandrasekhar-1964b}, and (c) the method  based on the dependence of the gravitational mass $M$ and the radius $R$ on the central energy density ${\cal E}_c$ (hereafter traditional method TM). The stability condition demands that the mass increases with increasing central energy density $dM/d{\cal E}_c>0$. The  extrema in the mass indicates a change in the stability of the compact star configuration~\cite{Shapiro-1983,Glendenning-2000,Haensel-2007,Zeldovich-71,Weinberg-72,Schutz-85,Bielich-2020}.  

In the present work we employ the third method in order to investigate a possible relation between the stability of a relativistic star and the corresponding CE. In this case, the questions that arise and must be answered (or at least investigated) are the following: Is there any  one-to-one correspondence between the minimum of the CE and the stability point for each realistic EoS? Is this  rule universal or it depends on the specific character of each EoS? Is it possible, even in some special cases, to associate stability with minimization of CE, and if so, which is the underlying reason? Obviously, one can appreciate the importance of discovering new ways, beyond the classical ones, to find stability conditions for compact objects.

In view of the above questions, the main motivation of the present work is to provide an extended examination of the statements of Refs.~\cite{Gleiser-2015a} and \cite{Koliogiannis-2023}. In  Ref.~\cite{Gleiser-2015a} the authors found  that the stability regions of neutron stars (in the framework  of the simple Fermi gas model)  as well as of self-gravitating configurations of complex scalar field (boson stars) with various self couplings (detailed reviews in Refs.~\cite{Liddle-1992,Liebling-2017,Schunck-2003}), obtained from traditional perturbation methods, correlate the critical points of the CE with an accuracy of a few percent. In Ref.~\cite{Koliogiannis-2023} the authors found that the suggested prediction of the stability by the minimization of the CE, concerning neutron stars and quark stars, does not have, at least quantitatively, universal validity although in several cases it qualitatively predicts the existence of the stability point. In this work, we aim to further elucidate this issue by seeking to reconcile these seemingly contradictory findings.  Specifically, we calculate the configurational entropy of bosonic and fermionic systems, described by interacting Fermi and boson gases, respectively, that form compact objects stabilized by gravity. We investigate whether the minimization of configurational entropy coincides with the stability point of the corresponding compact objects. Our results indicate a strong correlation between the stability points predicted by configurational entropy and those obtained through traditional methods, with the accuracy of this correlation showing a slight dependence on the interaction strength.   

The paper is organized as follows. In Sec.~\ref{sec:hydro_equi} we present the basic formalism of the hydrodynamic equilibrium and the role of analytical solutions  while in Sec.~\ref{sec:conf_entr}, we review the definition of the configurational entropy. The parametrization of the equations of state is provided in Sec.~\ref{sec:EoS} and and the results of the present study are laid out and discussed in Sec.~\ref{sec:results}. Finally, Sec.~\ref{sec:remarks} contains the concluding remarks.

\section{Hydrodynamic equilibrium and analytical solutions} \label{sec:hydro_equi}
To construct the related configuration in each compact object, which is the key property to calculate the CE, we employ the Einstein's field equations of a spherical fluid. In this case the mechanical equilibrium of the star matter is determined by the well known
Tolman-Oppenheimer-Volkoff (TOV) equations~\cite{Shapiro-1983,Glendenning-2000,Haensel-2007,Zeldovich-71}
\begin{eqnarray}
\frac{dP(r)}{dr}&=&-\frac{G{\cal E}(r) M(r)}{c^2r^2}\left(1+\frac{P(r)}{{\cal E}(r)}\right) \nonumber \\
&\times& \left(1+\frac{4\pi P(r) r^3}{M(r)c^2}\right) \left(1-\frac{2GM(r)}{c^2r}\right)^{-1},
\label{TOV-1}
\end{eqnarray}
\begin{equation}
\frac{dM(r)}{dr}=\frac{4\pi r^2}{c^2}{\cal E}(r).
\label{TOV-2}
\end{equation}

In general, to obtain realistic solutions, it is most natural to numerically solve the TOV equations by incorporating an EoS that describes the relationship between pressure and energy density within the fluid interior. Alternatively, one can seek analytical solutions to the TOV equations, though these solutions may lack physical relevance. While there are hundreds of analytical solutions to the TOV equations~\cite{Kramer-1980,Delgaty-1998}, only a few are of significant physical interest. In this work, we employ two of these noteworthy solutions: the Schwarzschild (constant-density interior solution) and the Tolman-VII solution~\cite{Kramer-1980,Delgaty-1998}. It is important to note that analytical solutions are highly valuable, as they often provide explicit expressions for the quantities of interest and are instrumental in verifying the accuracy of numerical calculations. Below, we briefly describe these two fundamental analytical solutions.

\begin{itemize}
    \item \textbf{Schwarzschild  solution}: In the case of the Schwarzschild interior solution, the density is constant throughout the star~\cite{Weinberg-72,Schutz-85}. The energy density and the pressure read as
    \begin{eqnarray}
        {\cal E}&=&{\cal E }_c=\frac{3M}{4\pi R^3}, \\
        \frac{P(x)}{{\cal E}_c }&=&\frac{\sqrt{1-2\beta}-\sqrt{1-2\beta x^2}}{\sqrt{1-2\beta x^2}-3\sqrt{1-2\beta }},
        \label{Unif-E}
    \end{eqnarray}
    where $x=r/R$, $\beta=GM/Rc^2$ is the compactness of the star and  ${\cal E}_{c}=\rho_c c^2$ is the central energy density. 

    \item \textbf{Tolman-VII solution}: The Tolman-VII solution has been extensively employed in neutron star studies while its physical realization has been examined, very recently, in detail~\cite{Oppenheimer-39,Raghoonundun-2015}. The stability of this solution has been examined by Negi {\it et al}.~\cite{Negi-1999,Negi-2001} and also confirmed in Ref.~\cite{Moustakidis-2017}. The energy density and the pressure read as~\cite{Moustakidis-2017}
    \begin{eqnarray}
        \frac{{\cal E} (x)}{{\cal E}_c}&=&(1-x^2), \quad {\cal E}_c=\frac{15Mc^2}{8\pi R^3},\\
        \frac{P(x)}{{\cal E}_c}&=&\frac{2}{15}\sqrt{\frac{3e^{-\lambda}}{\beta }}\tan\phi-\frac{1}{3}+\frac{x^2}{5}.
        \label{Tolm-E}
    \end{eqnarray}
\end{itemize}
In is worth mentioning that these solutions are applicable to any kind of compact object independently of the values of mass and radius. Consequently, they are suitable in studying any kind of massive or supramassive object which the hydrodynamic stability obeys to TOV equations. In any case, useful insight can be gained by the use of analytical solutions concerning both the qualitative and quantitative behavior of CE as a function of central density $\rho_c$.

\section{Configurational entropy in momentum  space } \label{sec:conf_entr}
The key quantity to calculate the CE in momentum space is the Fourier transform $F({\bf k})$ of the density $\rho(r)={\cal E}(r)/c^2$, originating from the solution of the TOV equations, that is
\begin{eqnarray}
F({\bf k})&=&\int\int\int \rho(r)e^{-i {\bf k}\cdot {\bf r}} d^3{\bf r} \nonumber \\ &=&\frac{4\pi}{k}\int_0^R \rho(r)r \sin(kr)dr.
\label{Fk-1}
\end{eqnarray}
It is  notable that the function $F({\bf k})$ in the case of zero momentum, coincides  with the gravitational mass of the compact object, that is $F(0)\equiv M$, since by definition (see Eq.~\eqref{TOV-2})
\begin{equation}
M=4\pi \int_0^R \rho(r) r^2 dr.
\label{Mas-1}
\end{equation}
Moreover, we define the modal fraction $f({\bf k})$~\cite{Gleiser-2015a}
\begin{equation}
f({\bf k})=\frac{|F({\bf k})|^2}{\int|F({\bf k})|^2d^3{\bf k}},
\label{Fk-2}
\end{equation}
and also the function $\tilde{f}({\bf k})=f({\bf k})/f({\bf k})_{\rm max}$, where $f({\bf k})_{\rm max}$ is the maximum fraction, which is given in many cases by the zero mode $k=0$, or by the system's longest physics mode, $|k_{\rm min}|=\pi/R$. The above normalization guarantees that $\tilde{f}({\bf k}) \leq 1$ for all values of ${\bf k}$. 

Finally, the CE, $S_C$, as a functional of $\tilde{f}({\bf k})$, is given by
\begin{equation}
S_C[\tilde{f}]=-\int \tilde{f}({\bf k}) \ln [\tilde{f}({\bf k})] d^3{\bf k}.
\label{S-1}
\end{equation}

Summarizing, for each EoS, an infinite number of configurations can appear, leading to the construction of $M-{ \rho}_c$ and $S_C$-${\rho}_c$ dependence. The latter fascilitates the investigation of any possible correlation between the minimum of $S_C$ and the stability point of compact objects under examination.

\section{Equation of state} \label{sec:EoS}
The present study focuses on the role of the CE as a stability condition of compact objects obeying Fermi or boson statistics including a parametrized self-interaction. The compact objects under consideration are neutron stars (introduced by a simplified EoS), boson stars and other type of astrophysical objects composed of fermions or bosons, such as dark matter stars. In following, the equations of interacting Fermi and boson gases are introduced.
\subsection{Interacting Fermi gas (FG)}
\label{sec:fermi_gas}
In the case of compact objects consisting of solely interacting Fermi gas (FG), we considered the simplest extension of the free fermion gas in which an extra term that introduces the repulsive interaction between fermions is added. Therefore, the energy density and pressure of the fermions are described as (for an extensive analysis see Ref.~\cite{Narain-06}) 
~
\begin{eqnarray}
{\cal E}(n_{\chi})&=&\frac{(m_{\chi}c^2)^4}{(\hbar c)^38\pi^2}\left[x\sqrt{1+x^2}(1+2x^2) \right. \nonumber\\
&-&\left. \ln(x+\sqrt{1+x^2})\right] + \frac{y^2}{2} (\hbar c)^3 n_{\chi}^2, \\
P(n_{\chi})&=&\frac{(m_{\chi}c^2)^4}{(\hbar c)^38\pi^2}\left[x\sqrt{1+x^2}(2x^2/3-1)
\right. \nonumber \\
&+&\left. \ln(x+\sqrt{1+x^2})\right]  + \frac{y^2}{2} (\hbar c)^3 n_{\chi}^2,
\label{Rel-ED}
\end{eqnarray}
where $m_{\chi}$ is the particle mass, considered equal to $m_{\chi}=939.565~{\rm MeV/c^{2}}$ for reasons of simplicity (this holds throughout the study), $n_{\chi}$ is the number density  and 
\[x=\frac{(\hbar c)(3\pi^2n_{\chi})^{1/3}}{m_{\chi}c^2}.  \]
The last parameter, $y$ (in units of $\rm MeV^{-1}$), is the one that introduces the repulsive interaction. In this study we have considered the values $y=[0,0.001,0.005,0.01,0.05,0.1,0.3,0.5]~(\rm MeV^{-1})$ where increasing $y$ increases the strength of the interaction and vice-versa.

\subsection{Interacting boson gas (BG)}
\label{sec:boson_gas}
To enrich our study, we included the case of compact stars composed of bosonic matter and specifically that of an interacting boson gas. As the construction of the EoS for the aforementioned gas is not unambiguously defined, we introduce three cases based on different assumptions. It needs to be noted that since the scalar field only vanishes at spatial infinity, boson stars do not have a specific radius where the energy density and pressure vanish. Thus, we do not use a momentum cut-off scheme, $0\leq |{\bf k}|\leq\infty$. 

\begin{enumerate}
    \item \textbf{BG-C1}: The first derivation of the EoS of boson stars with a repulsive interaction was given in  Refs.~\cite{Colpi-1986,PhysRevD.105.023001,PhysRevD.109.043029} and since then it has been used extensively in the corresponding calculations. In particular, the energy density is given as
    \begin{equation}
    {\cal E}(P)= \frac{4}{3w} \left[ \left(\sqrt{\frac{9w}{4}P}+1  \right)^2-1 \right],
    \label{eq:boson_matter_a}
    \end{equation}
    where $w=4\lambda (\hbar c)^3/(m_{\chi}c^2)^4$ (in units of $\rm MeV^{-1}~fm^{3}$). In fact, the parameter $\lambda$ is the one that is related with the strength of the interaction. However, it is usual to employ the combination of $m_{\chi}$ and $\lambda$ defined as $w$. In this study we have considered the values $w=[0.001,0.005,0.01,0.05,0.1,0.3,0.5]~(\rm MeV^{-1}~fm^{3})$ where increasing $w$ increases the strength of the interaction and vice-versa.

    \item \textbf{BG-C2}: The second way to describe the interior of a boson star is through the EoS provided in Ref.~\cite{Agnihotri-2009} and used recently in Ref.~\cite{Watts-2023}, where the energy density is given by
    \begin{equation}
    {\cal E}(P)=P +\sqrt{\frac{2P}{z}},
    \label{eq:boson_matter_b}
    \end{equation}
     with the interaction parameter $z$ being equal to $z=u^2(\hbar c)^3/(m_{\chi} c^2)^2$ (in units of $\rm MeV^{-1}~fm^{3}$) and the quantity $u={\rm g}_{\chi}/m_{\phi}c^2$ defines the strength of the interaction in analogy to the case of fermions. In this study we have considered the values $z=[0.001,0.005,0.01,0.05,0.1,0.3,0.5]~(\rm MeV^{-1}~fm^{3})$.

    \item \textbf{BG-C3}: Recently, in Ref.~\cite{Pitz-2023} the authors studied the properties of self-interacting boson stars with different scalar potentials. They concluded that the resulting properties of the boson star configurations differ considerably from previous calculations. Therefore, to enhance the connection between the stability criterion and the CE, we employed two cases of the EoSs introduced in Ref.~\cite{Pitz-2023}: (a) one with a mass term (MT) and (b) one with a vacuum term (VT) without a mass term. The scaling EoSs read as
    \begin{equation}
        \mathcal{E}(P) = \begin{cases}
        P^{2/n}+(n+2)P/(n-2), & \text{MT}, \\
        1+(n+2)P/(n-2), & \text{VT},
        \end{cases}
    \end{equation}
    where the index $n$ is restricted to $n>2$. In this study we have considered the values $n=[4,5]$ for MT [hereafter case (a)] and $n=[3,4,5]$ for VT [hereafter case (b)]. An additional reason for using the mentioned cases is because they lead to different mass-radius diagrams (depending on the index values) and thus, cover a large range of cases that may correspond to boson stars.

\end{enumerate}

\noindent We consider that pluralism in the use of EoS will greatly help to test the plausibility of the stability criterion through the CE in the case of boson stars.

\begin{figure*}
\includegraphics[width=0.9\textwidth]{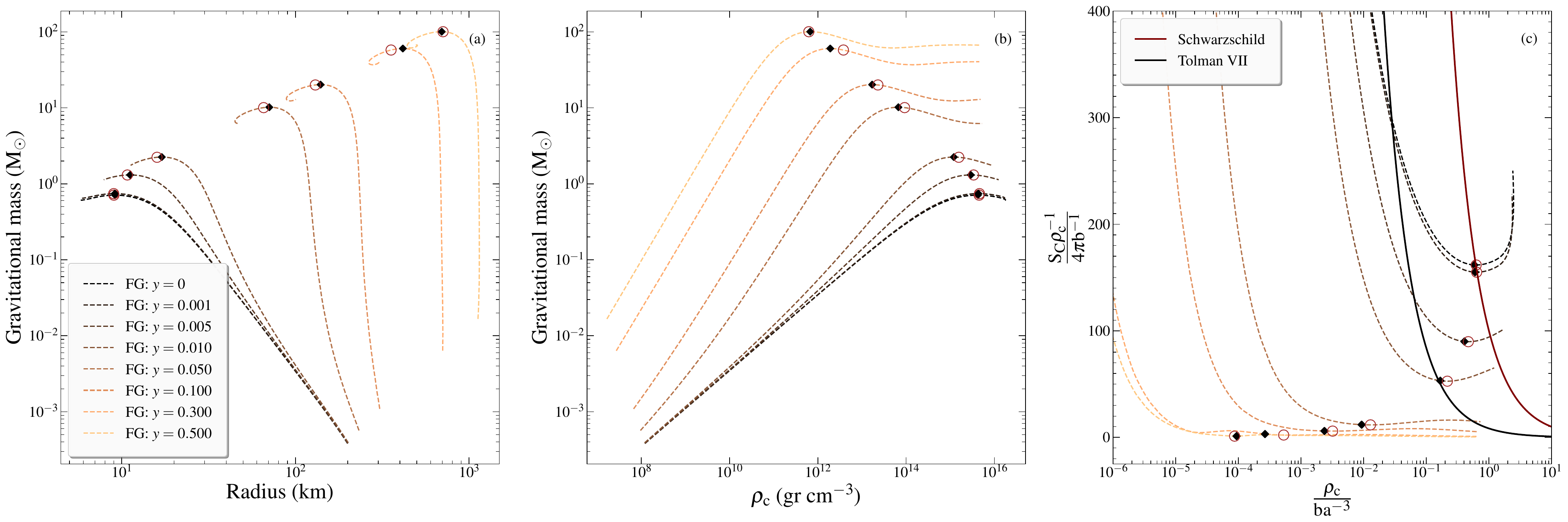}
\includegraphics[width=0.9\textwidth]{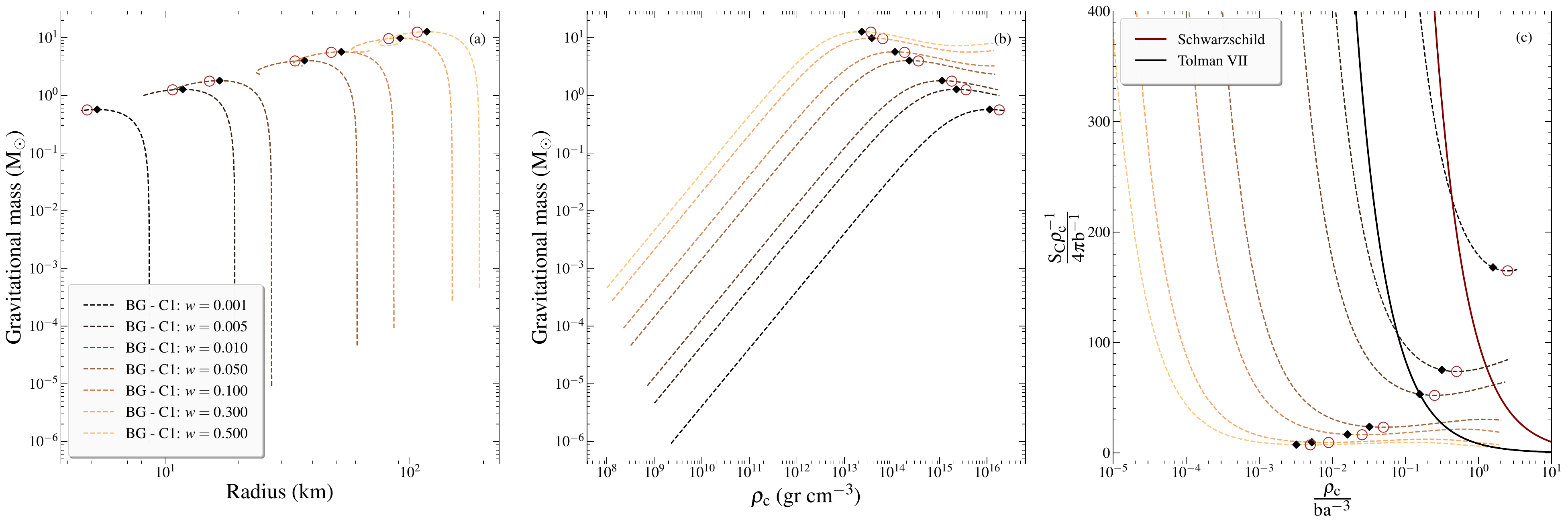}
\includegraphics[width=0.9\textwidth]{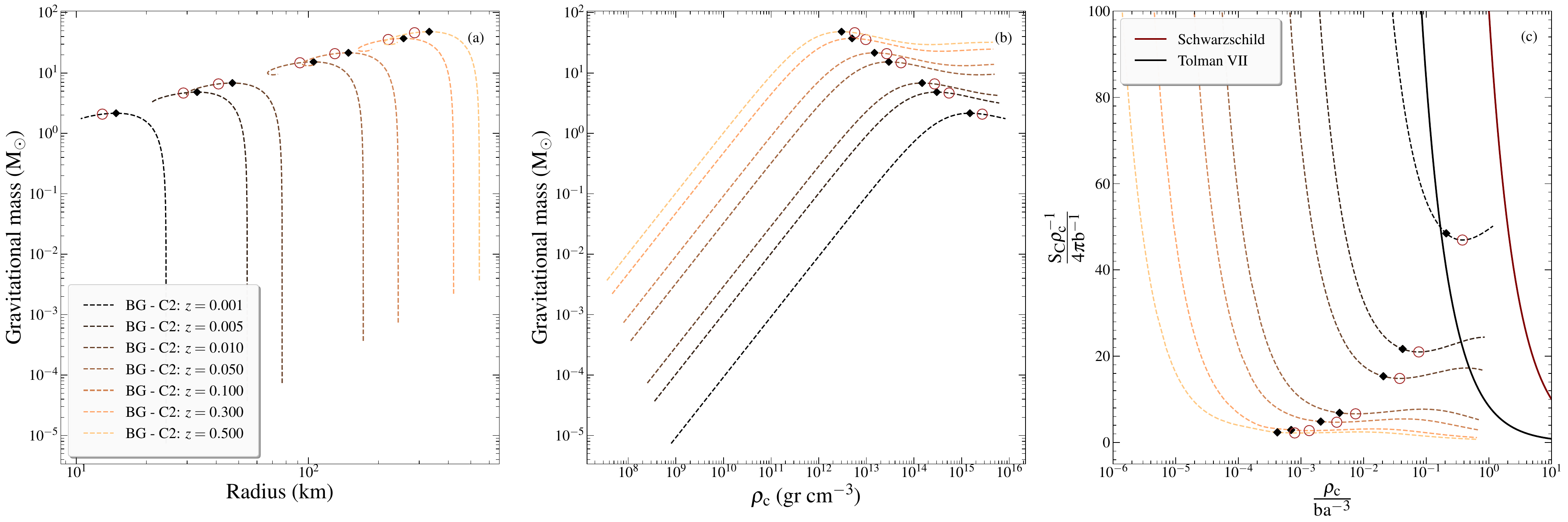}
\includegraphics[width=0.9\textwidth]{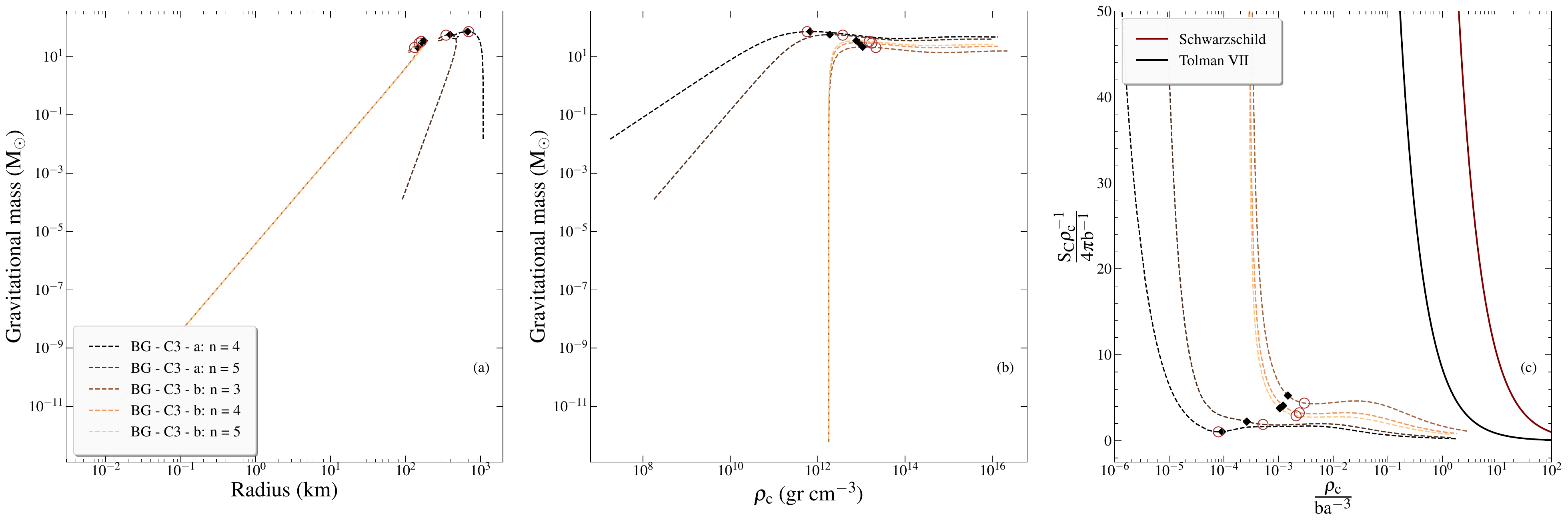}
\caption{(a) Gravitational mass as a function of the radius. (b) The corresponding dependence of the gravitational mass as a function of the central density. (c) The corresponding CE as a function of the central density and two analytical solutions of TOV equations, where their constants $\rm a$ and $\rm b$ are given in the text below Eq.~\eqref{eq:analytical_sol} and ensure the dimensionless units of the two axes. The black diamonds indicate the stability points due to the TM while the open circles correspond to the minimum of the CE. The panels in order correspond to FG EoSs, BG-C1 EoSs, BG-C2 EoSs, BG-C3 EoSs.
}
\label{fig:mr_entropy_plot}
\end{figure*}

\section{Results and Discussion} \label{sec:results}
As a first step, we employed two analytical solutions, namely the Schwarzchild's and Tolman's-VII solutions, in order to calculate the CE. It is worth pointing out that although it is more natural to use a realistic EoS for the fluid interior in order to solve the Einstein’s field equations, the use of analytical solutions has the advantage that by having an explicit form, the examination of the implied physics becomes simpler. We should remark that the analytical solutions are a source of infinite number of EoSs (plausible or not). Consequently, they can be used extensively, to introduce and establish some universal approximations (for more details and discussion see also Ref.~\cite{Moustakidis-2017}). In both cases, we formulate the dependence in the form
\begin{equation}
  \frac{S_C \rho_{c}^{-1}}{4\pi b^{-1}}={\cal C}\times 10^5\times \left(\frac{\rm km}{R}  \right)^3\frac{1}{\rho_c/ba^{-3}},
  \label{eq:analytical_sol}
\end{equation}
where
\[a=\frac{1}{\pi}\left(\frac{h}{mc}  \right)^{3/2}\frac{c}{(mG)^{1/2}}, \quad b=\frac{c^2}{G}a,  \] and ${\cal C}$ taking the values 1.728  and 0.145 for the Schwarzschild and Tolman-VII solutions, respectively.
  
Although the above expression does not ensure the location of the stability point, it is very useful for two reasons: (a) comparison with the results produced by using realistic EoSs, for a fixed value of the radius $R$ (see Fig.~\ref{fig:mr_entropy_plot}\textcolor{blue}{(c)}), and (b) check and ensure the accuracy of our numerical calculations.

In Fig.~\ref{fig:mr_entropy_plot} we display in order the four cases corresponding to: (first) Fermi gas, (second) boson gas - C1, (third) boson gas - C2, and (fourth) boson gas - C3. In particular, Fig.~\ref{fig:mr_entropy_plot}\textcolor{blue}{(a)} manifests the dependence of the gravitational mass on the radius, Fig.~\ref{fig:mr_entropy_plot}\textcolor{blue}{(b)} presents the dependence of the gravitational mass on the central density, and Fig.~\ref{fig:mr_entropy_plot}\textcolor{blue}{(c)} indicates the CE as a function of the central density for various values of the interaction parameters. In addition, diamonds demonstrate the stability points due to the TM, while open circles mark the minimum of the CE.

In the case of the FG, the first panel of Fig.~\ref{fig:mr_entropy_plot} displays that the points due to TM and CE are located in close proximity, validating the CE method for the location of the stability point. However, a detailed presentation of the percentage error on some interesting quantities, namely the gravitational mass, the radius, the central density and the compactness, as shown in Table~\ref{tab:table1}, signal a different behavior. While the error in the gravitational mass is lower than 5\%, the error in the central density can reach up to almost 99\%, depending each time to the value of the interaction. The latter have its origin in the creation of a plateau immediately after the rapid decrease of the CE. The existence of a plateau maintains the CE in a narrow region, while the central density is spanning in a wide region. In addition, as the data indicate, there is no simple relation between the interaction and the corresponding error to establish a pattern. Thus, in the FG case, while the CE can potentially establish the maximum gravitational mass with good accuracy, the proper description of the central density is almost impossible. It needs to be noted that for some specific values of the interaction, the location of the total minimum in CE was not successful, even at high values of densities beyond the maximum mass configuration (unstable region). In that cases, we located a local minimum near the density that corresponds to the maximum mass configuration. The aforementioned statement holds for all cases under consideration in the present study.

\begin{table}[b!]
	\caption{Percentage $(\%)$ error in gravitational mass, radius, central density and compactness of the CE method with respect to the TM. The interaction $\rm y$ is in units of $\rm MeV^{-1}$, while $\rm w$ and $\rm z$ in $\rm MeV^{-1}~fm^{3}$.}
	\begin{ruledtabular}
		\begin{tabular}{l r S r r r r}
			& & {$\rm (y,w,z,n)$} & \multicolumn{1}{c}{$\rm M$} & \multicolumn{1}{c}{$\rm R $} & \multicolumn{1}{c}{$\rm \rho_{c}$} & \multicolumn{1}{c}{$\rm \beta$} \\
			\hline
			\multirow[t]{8}{*}{FG (y)} & & 0.000 & 0.030 & 1.648 & 7.034 & 1.645 \\
            & & 0.001 & 0.022 & 1.442 & 6.014 & 1.440 \\
            & & 0.005 & 0.176 & 3.580 & 15.320 & 3.530 \\
            & & 0.010 & 0.655 & 6.158 & 29.650 & 5.863 \\
            & & 0.050 & 1.177 & 7.276 & 38.295 & 6.577 \\
            & & 0.100 & 0.984 & 6.743 & 35.102 & 6.176 \\
            & & 0.300 & 4.433 & 14.491 & 98.601 & 11.762 \\
            & & 0.500 & 0.093 & 1.690 & 7.006 & 1.754 \\
            \multicolumn{7}{c}{\vspace{-0.25cm}} \\
            \multirow[t]{7}{*}{BG-C1 (w)} & & 0.001 & 1.776 & 9.012 & 58.881 & 7.953 \\
            & & 0.005 & 1.789 & 9.045 & 59.167 & 7.978 \\
            & & 0.010 & 1.789 & 9.120 & 59.813 & 8.067 \\
            & & 0.050 & 1.753 & 8.591 & 55.327 & 7.480 \\
            & & 0.100 & 1.916 & 9.001 & 58.827 & 7.785 \\
            & & 0.300 & 2.252 & 10.222 & 69.804 & 8.877 \\
            & & 0.500 & 1.750 & 8.585 & 55.422 & 7.477 \\
            \multicolumn{7}{c}{\vspace{-0.25cm}} \\
            \multirow[t]{7}{*}{BG-C2 (z)} & & 0.001 & 3.396 & 12.756 & 81.381 & 10.728 \\
            & & 0.005	& 3.436 & 12.709 & 80.985 & 10.622 \\
            & & 0.010	& 3.369 & 12.891 & 82.586 & 10.931 \\
            & & 0.050	& 3.237 & 12.633 & 80.190 & 10.755 \\
            & & 0.100	& 3.238 & 12.627 & 80.190 & 10.746 \\
            & & 0.300	& 4.161 & 14.080 & 94.215 & 11.544 \\
            & & 0.500	& 3.904 & 13.557 & 89.078 & 11.167 \\
            \multicolumn{7}{c}{\vspace{-0.25cm}} \\
            \multirow[t]{5}{*}{BG-C3 (n)} & \multirow[t]{2}{*}{case (a)} & 4 & 0.418 & 3.195 & 14.332 & 3.501 \\
            & & 5 & 4.264 & 11.592 & 98.208 & 8.289 \\
            & \multirow[t]{3}{*}{case (b)} & 3 & 4.318 & 7.901 & 100.000 & 3.890 \\
            & & 4 & 4.822 & 7.957 & 97.959 & 3.406 \\
            & & 5 & 4.664 & 7.899 & 98.438 & 3.513

		\end{tabular}
	\end{ruledtabular}
	\label{tab:table1}
\end{table}

In the case of the BG, a similar behavior with the FG is observed. Once more, Fig.~\ref{fig:mr_entropy_plot} in the second, third and fourth panel, displays a visually small difference between the TM and CE points. Nevertheless, Table~\ref{tab:table1} illustrates that the error in the underlying quantities aligns with the trend observed in the FG. Specifically, the error in gravitational mass is less than 5\%, while the error in central density can extend up to 100\%. In this instance as well, the CE can be employed to determine the maximum mass, but not the associated central density. The observed behavior in both the FG and BG cases strengthens the argument that the location of the stability point is an intrinsic property of the EoS.

\begin{figure}
\includegraphics[width=0.703\columnwidth]{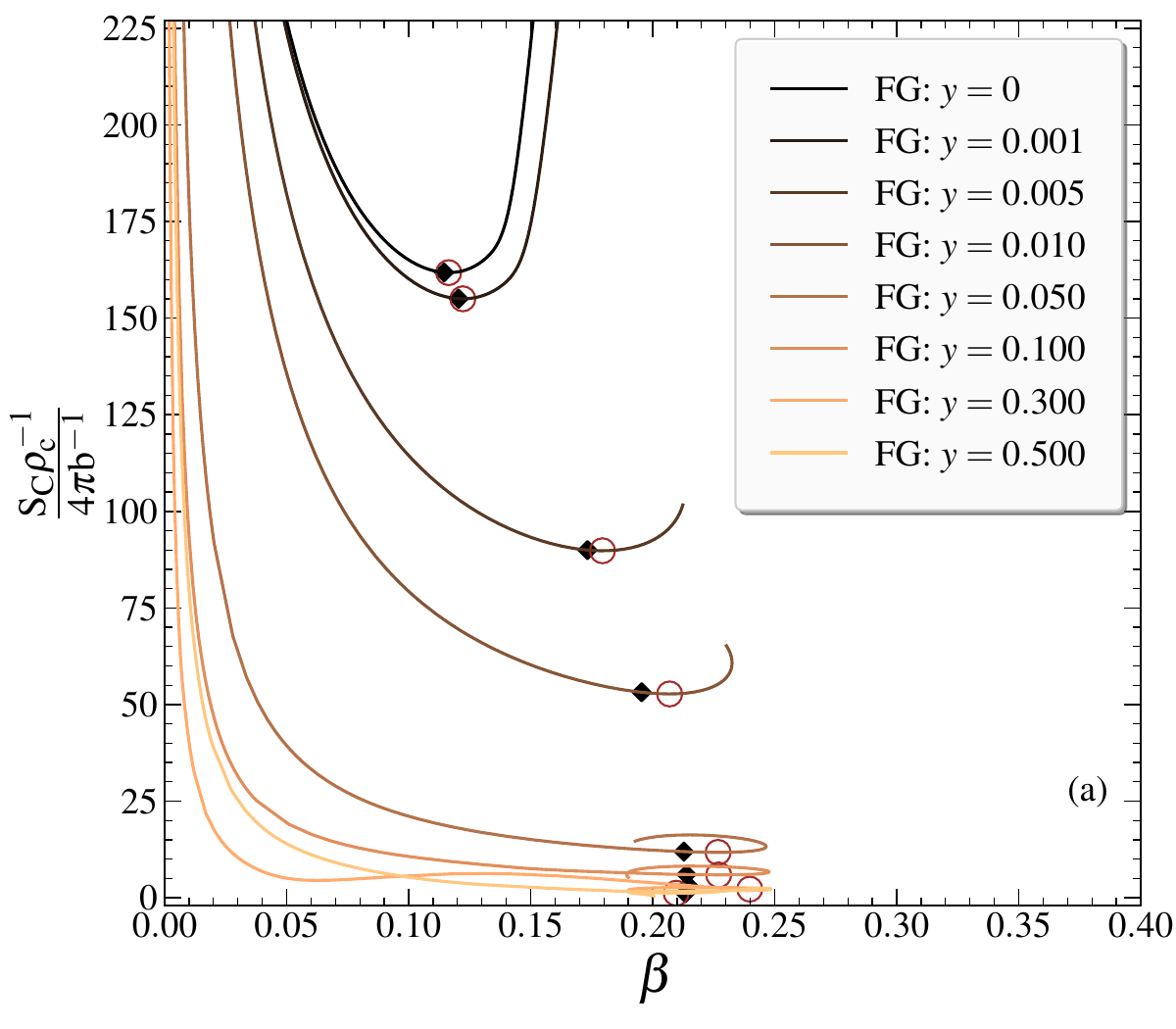}
~
\includegraphics[width=0.703\columnwidth]{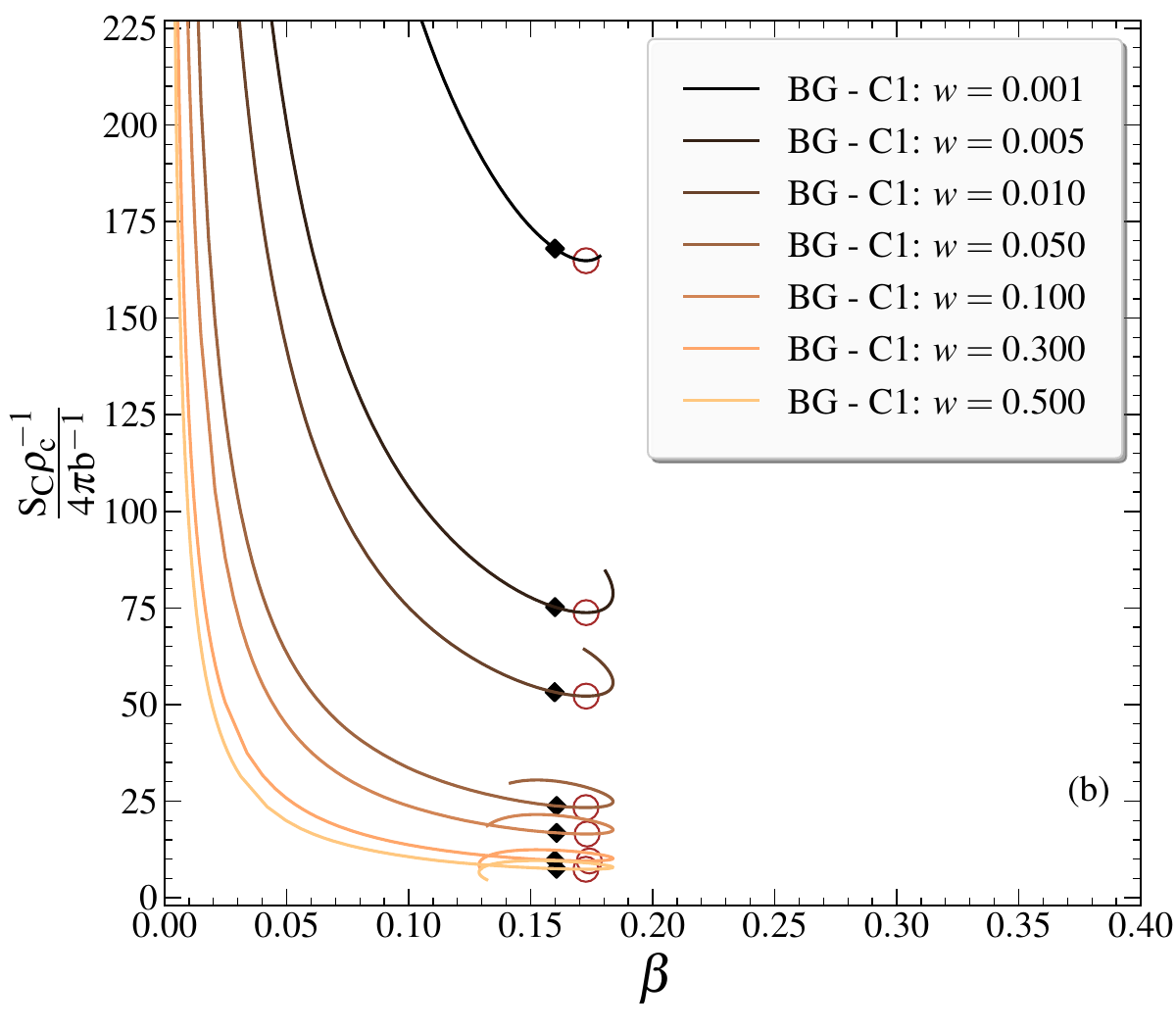}
~
\includegraphics[width=0.703\columnwidth]{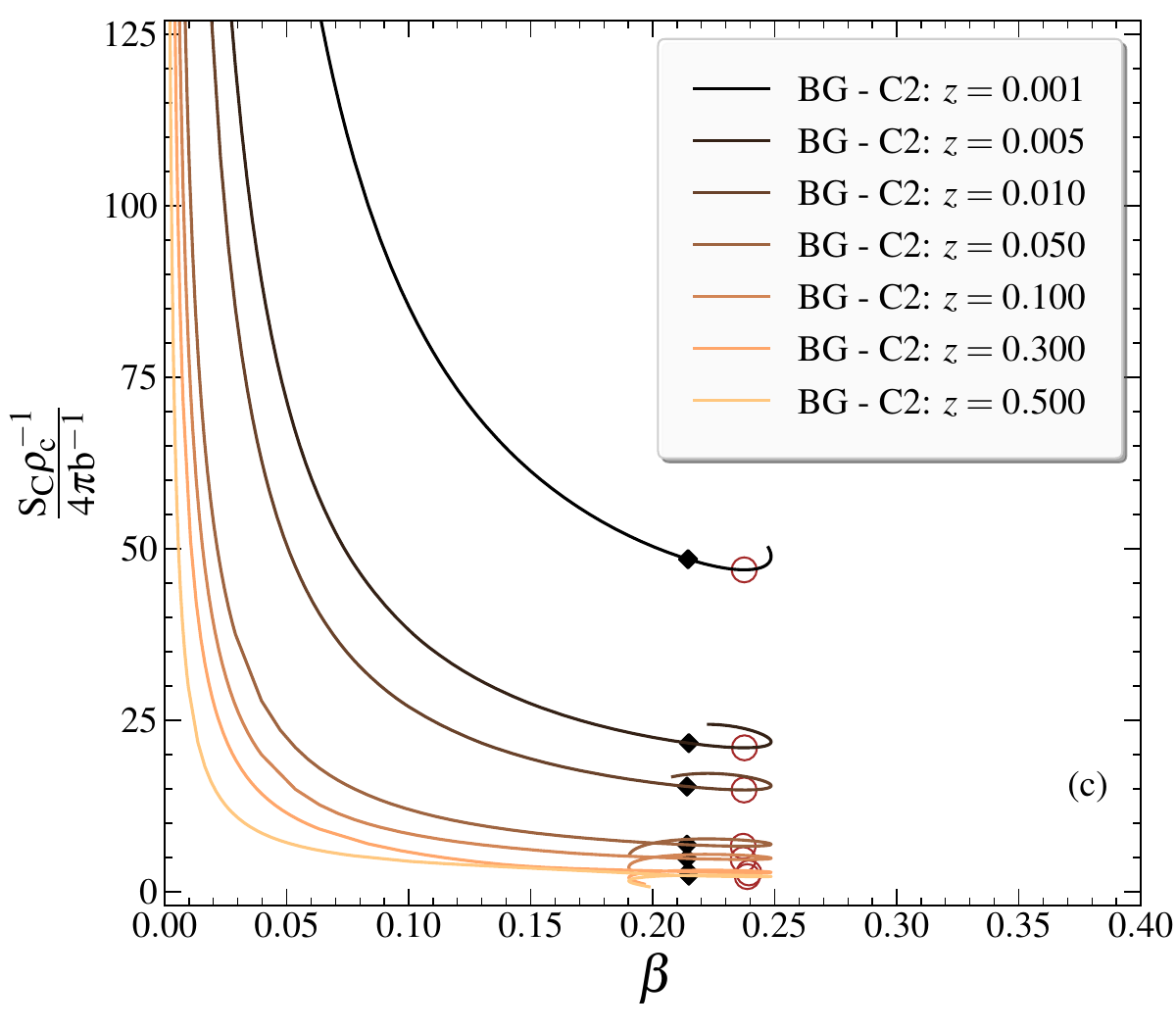}
~
\includegraphics[width=0.703\columnwidth]{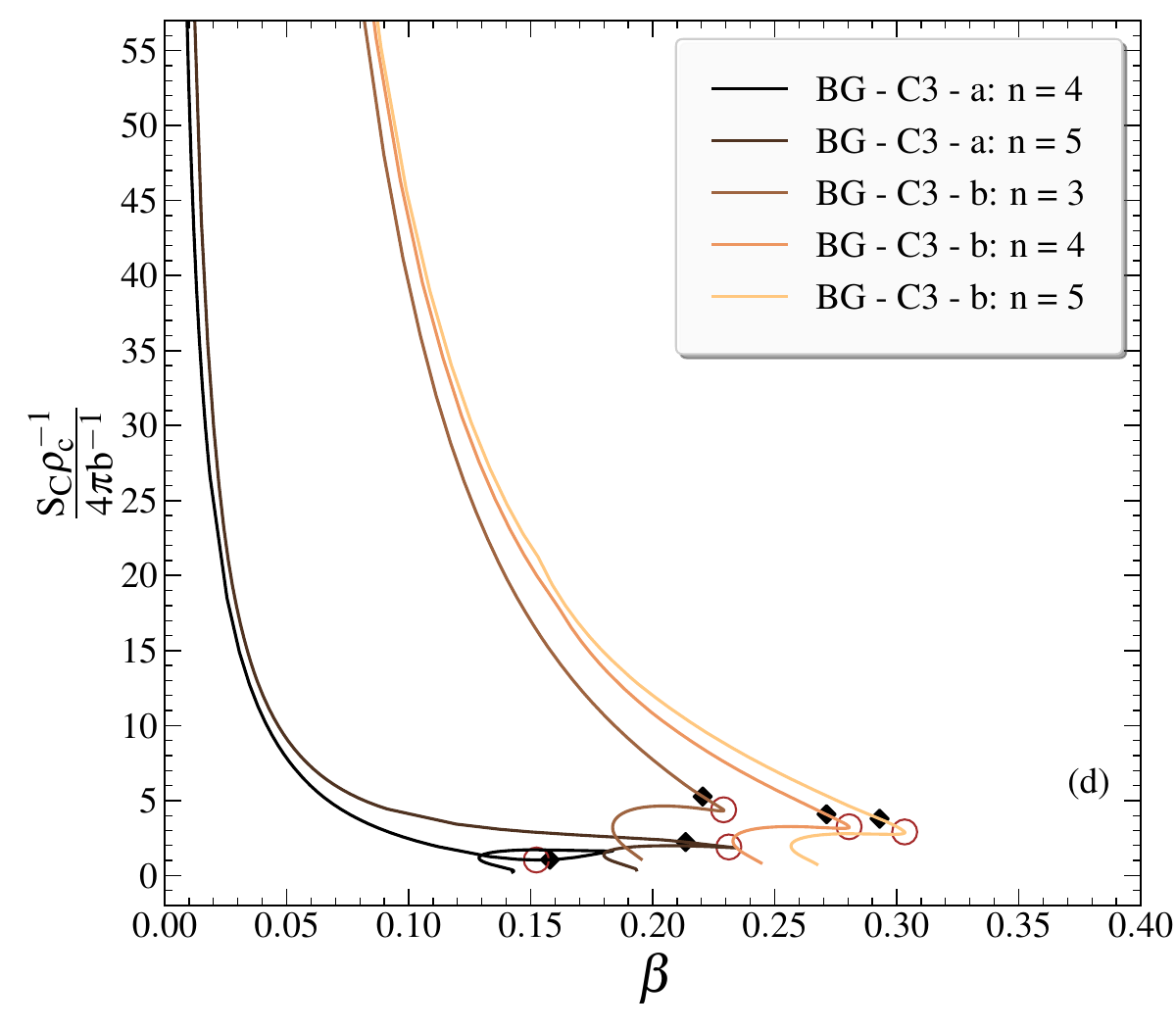}
\caption{ Configurational entropy as a function of the compactness parameter. The black diamonds indicate the stability points due to the TM while the open circles correspond to the minimum of the CE. The panels in order correspond to (a) FG EoSs, (b) BG-C1 EoSs, (c) BG-C2 EoSs, (d) BG-C3 EoSs.}
\label{fig:compactness}
\end{figure}

As anticipated, the CE is related to the star's compactness as illustrated in Fig.~\ref{fig:compactness}, which displays the dependence of the CE on the compactness for (a) FG, (b) BG-C1, (c) BG-C2, and (d) BG-C3. In the compactness plane the CE creates also a plateau, similar to the central density plane in Fig~.\ref{fig:mr_entropy_plot}\textcolor{blue}{(c)}, but in the majority of the cases it is not so extended, depending each time on the corresponding mass-radius diagram of Fig.\ref{fig:mr_entropy_plot}\textcolor{blue}{(a)}. In the FG case, the error can reach values up to 12\%, where in the majority of the cases the error is lower than 6\%. This result is in accordance with the error in the gravitational mass along with the corresponding error values in the radius. As the radius is more sensitive to the structure of the star, this sensitivity is also presented in the error, reaching values close to 15\%. Concerning the BG case, the error in the compactness is established in general under 12\% and in the corresponding radius under 14\%.

\begin{figure}
\includegraphics[width=0.48\columnwidth]{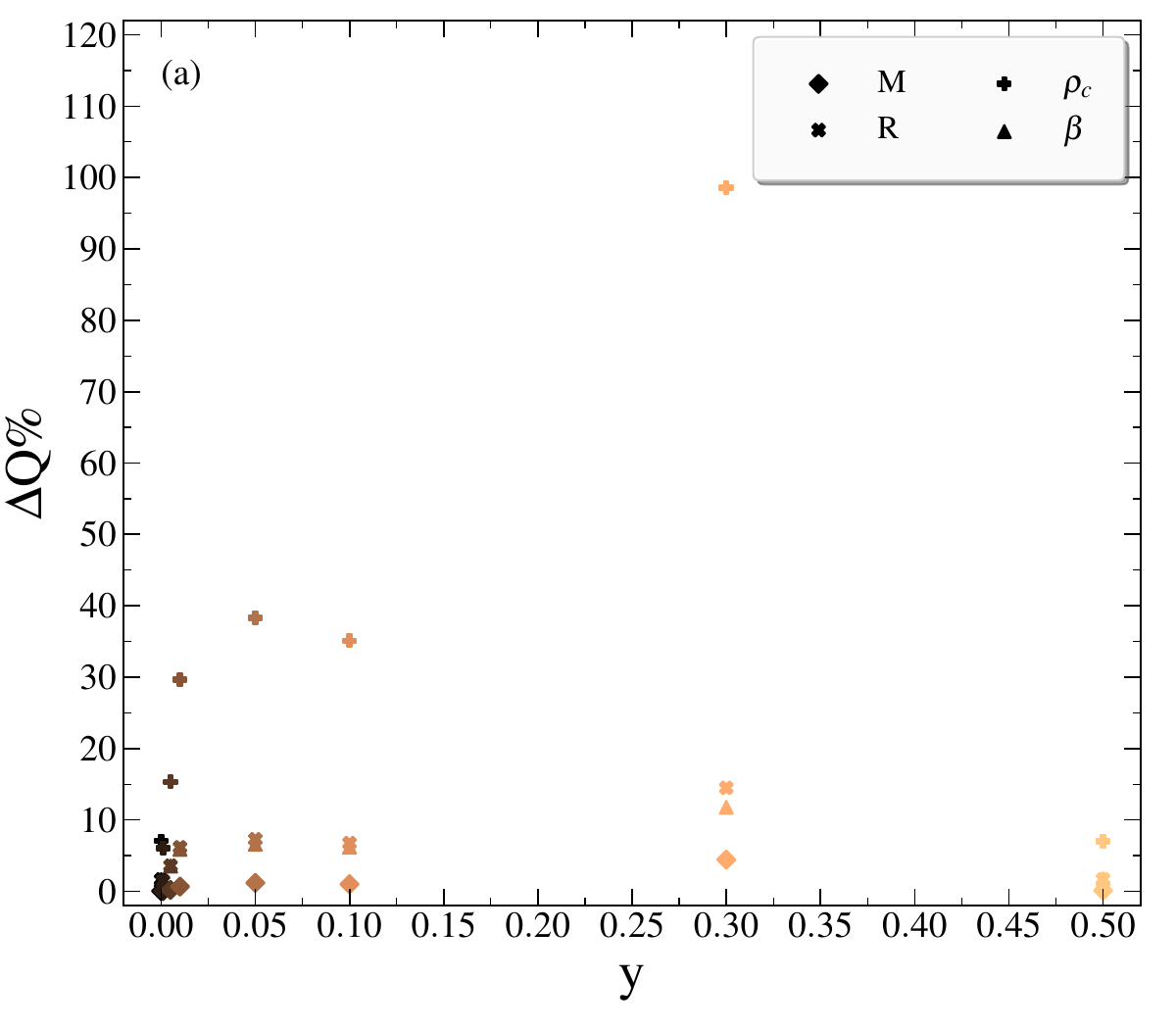}
~
\includegraphics[width=0.48\columnwidth]{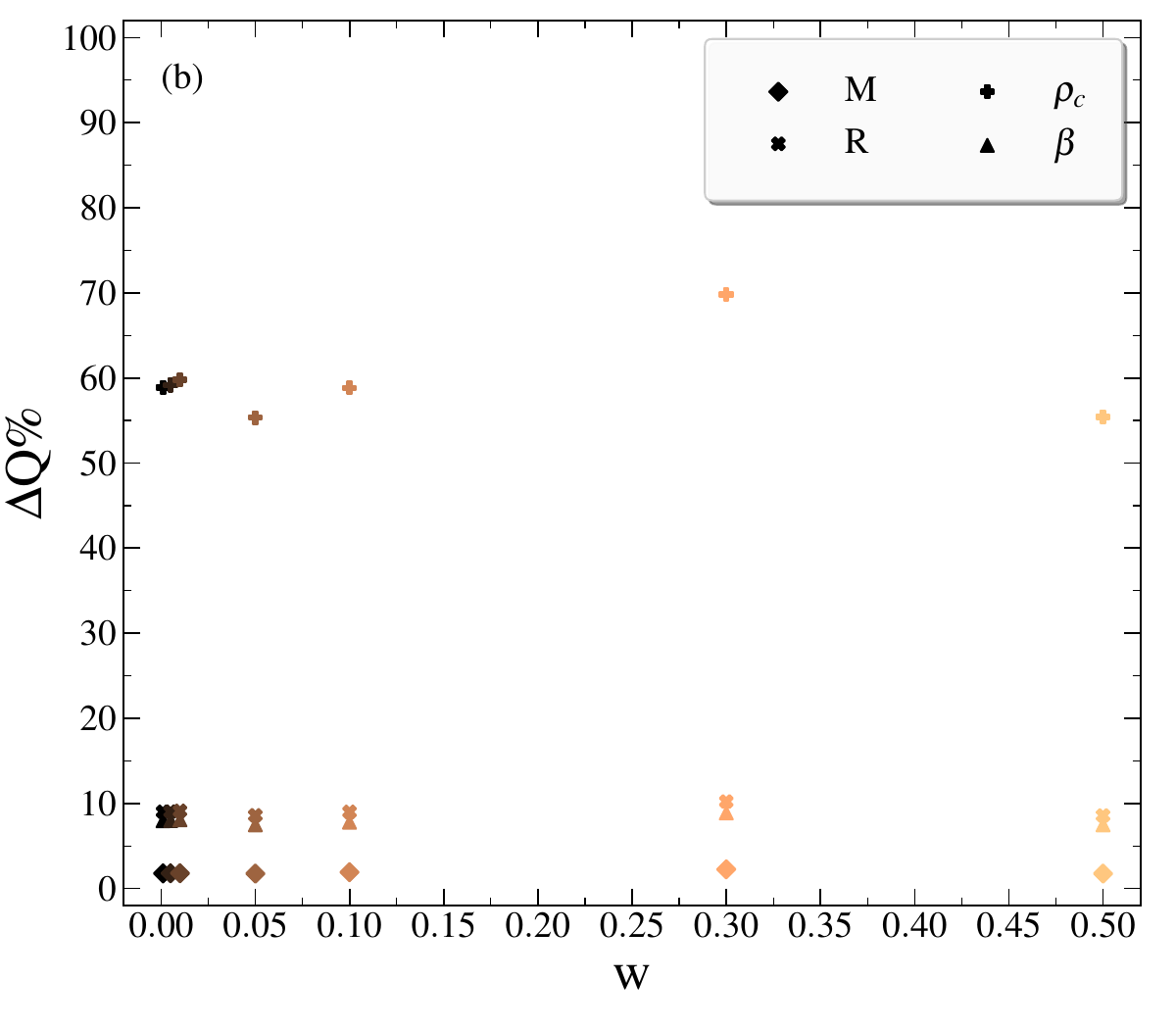}
~
\includegraphics[width=0.48\columnwidth]{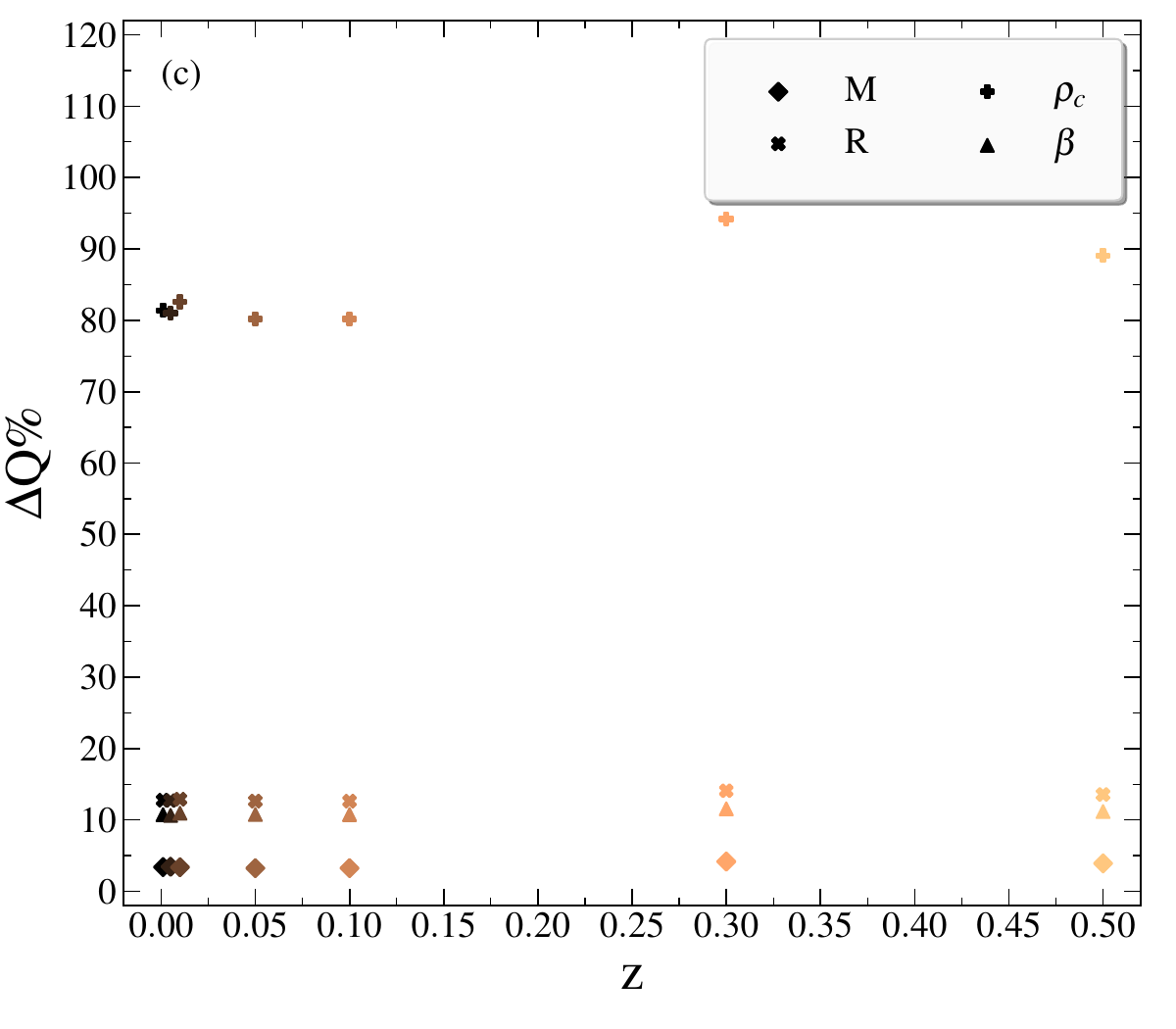}
~
\includegraphics[width=0.48\columnwidth]{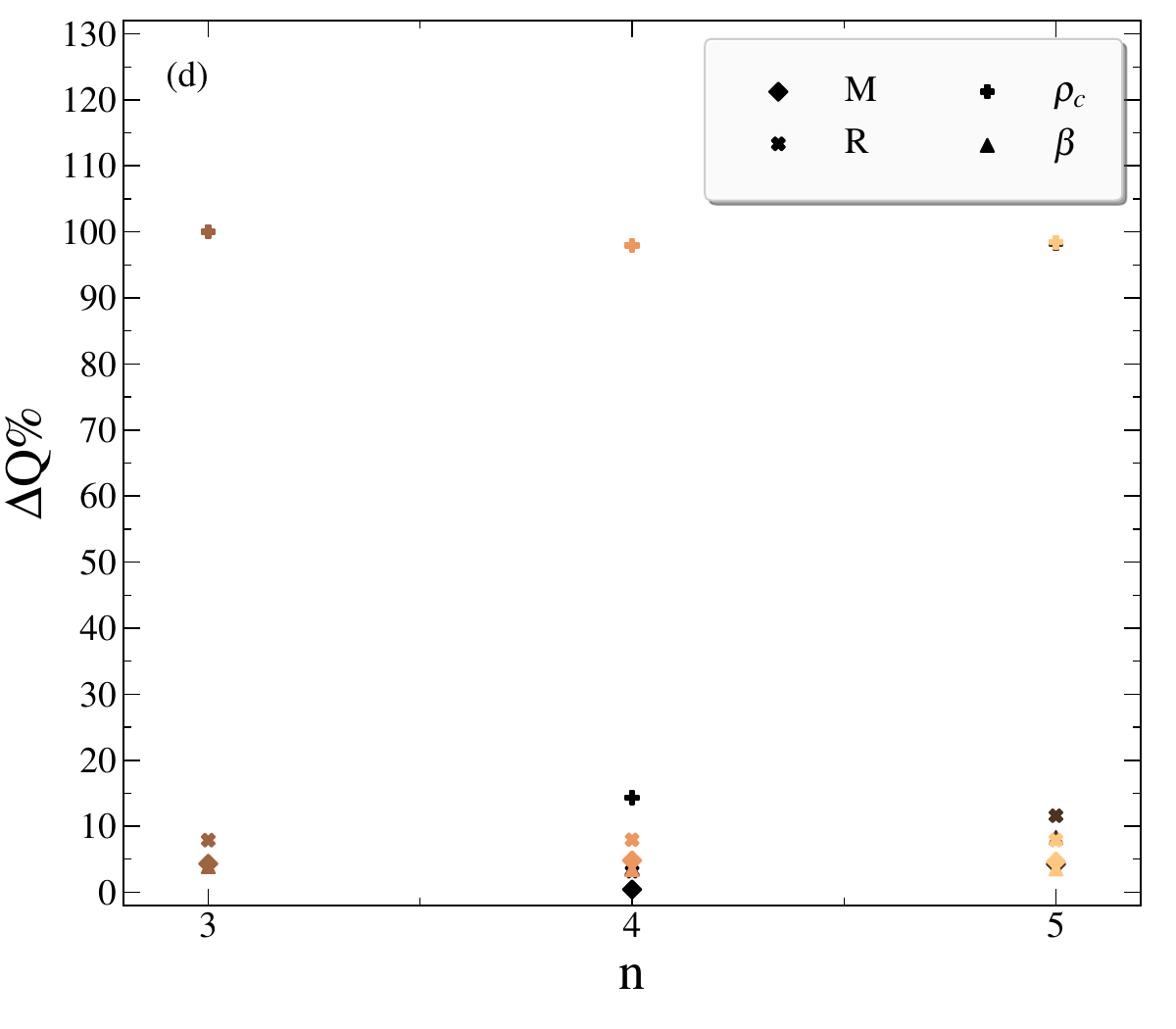}
\caption{Percentage error in gravitational mass (diamonds), radius (squares), central density (plus signs) and compactness (triangles) as a function of the interaction strength parameter of the CE method with respect to the TM method. The panels in order correspond to: (a) FG EoSs, (b) BG-C1 EoSs, (c) BG-C2 EoSs, (d) BG-C3 EoSs.}
\label{fig:errors}
\end{figure}

For a visual presentation of Table~\ref{tab:table1} for the predictions of the two methods, Fig.~\ref{fig:errors} displays the percentage error on the predictions for various fermionic and bosonic stars properties (gravitational mass, radius, central density and compactness) as  a function of the relative parametrization of the interaction strength $(y,w,z,n)$ in each particular case. As a general comment, the convergence of the two methods is clearly better in the case of fermion stars compared to that of boson stars (at least for the selected range of parametrizations). Furthermore, for the case of a boson star, the choice of the EoS is decisive for the accuracy of the convergence of the two methods.

\begin{figure}
\includegraphics[width=0.65\columnwidth]{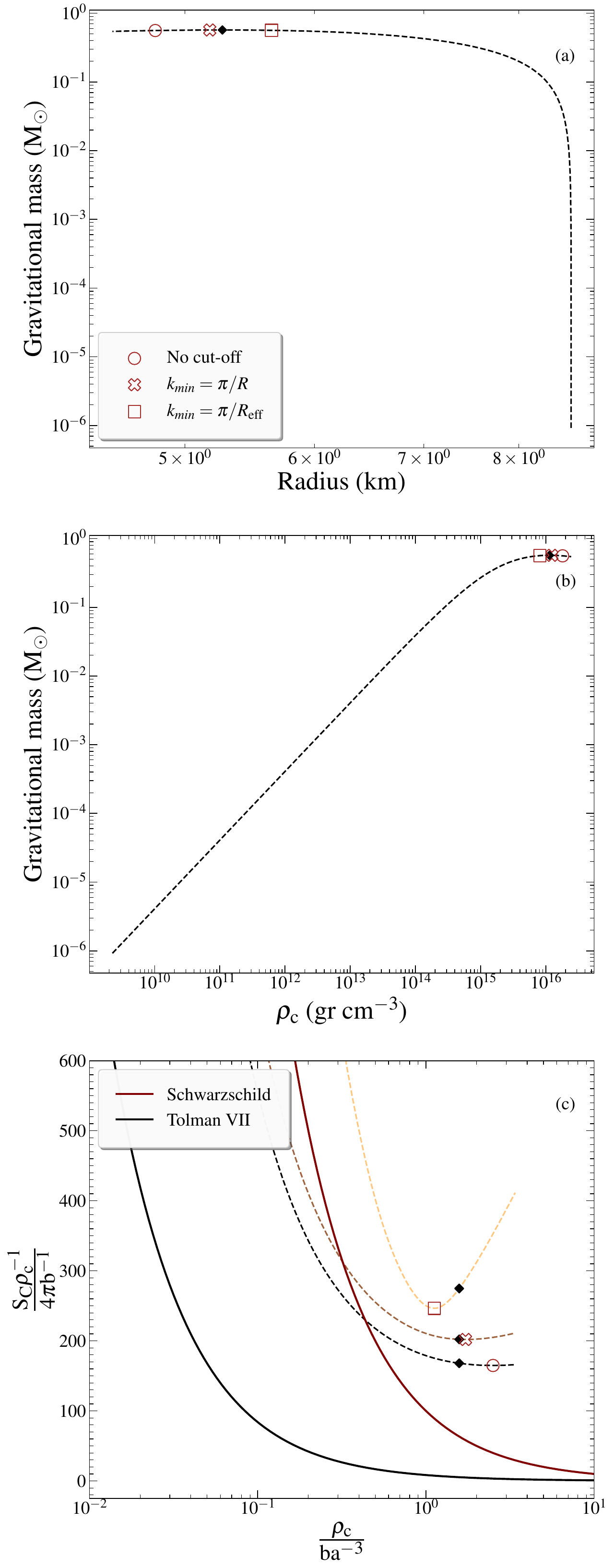}
\caption{(a) Gravitational mass as a function of the radius for the BG-C1 (0.001) EoS. (b) The corresponding dependence of the gravitational mass as a function of the central density. (c) The corresponding CE as a function of the central density and two analytical solutions of TOV equations, where their constants $\rm a$ and $\rm b$ are given in the text below Eq.~\eqref{eq:analytical_sol} and ensure the dimensionless units of the two axes. The black diamonds indicate the stability points due to the TM while the open markers correspond to the minimum of the CE with: (a) no cut-off ($k_{\rm min}=0$; circle), (b) cut-off at $k_{\rm min}=\pi/R$ (cross), and (c) cut-off at $k_{\rm min}=\pi/R_{\rm eff}$ (square).
}
\label{fig:cutoff_comp}
\end{figure}

As it is evident, one of the primary distinctions between fermionic and bosonic systems lies in the choice of the momentum cut-off scheme. In this context, we present a detailed analysis of its role for the systems under consideration. Specifically, we examine two cut-off schemes: (a) $k_{\rm min}=\pi/R$, which is based on the radius of the boson star originated from the condition that the pressure vanishes at the star's surface, $P(R)=0$, and (b) $k_{\rm min}=\pi/R_{\rm eff}$, which is the effective radius where most of the star's mass is concentrated and defined as~\cite{Gleiser-2015a}
\begin{equation}
    R_{\rm eff}\equiv\frac{\int_0^{\infty} \rho(r) r^3 dr}{\int_0^{\infty} \rho(r) r^2 dr}.
\end{equation}
The two schemes, along with the initial case where $k_{\rm min}=0$, are compared in detail in Fig.~\ref{fig:cutoff_comp}, for the BG-C1 (0.001) EoS as a representative case. The introduction of a cut-off scheme significantly improves the accuracy of the results, with case (a) showing noteworthy improvement in accuracy over the no cut-off case. Furthermore, the cut-off scheme in case (b) alters the behavior of the CE, leading to a sharper and more distinct outcome. The comparative results from the different cut-off schemes are also summarized in Table~\ref{tab:table2}, demonstrating that the choice of the momentum cut-off contributes significantly to the convergence of the two methods under consideration.

In conclusion, beyond the importance of the EoS, the selection of the appropriate cut-off scheme in the momentum plays a critical role in improving the accuracy and convergence of the two methods.

\begin{table}[H]
	\caption{Percentage $(\%)$ error in gravitational mass, radius, central density and compactness of the CE method with respect to the TM and various cut-off schemes for the BG-C1 (0.001) EoS. The interaction $\rm w$ is in units of $\rm MeV^{-1}~fm^{3}$.}
	\begin{ruledtabular}
		\begin{tabular}{l c r r r r r}
			& & \multicolumn{1}{c}{$\rm w$} & \multicolumn{1}{c}{$\rm M$} & \multicolumn{1}{c}{$\rm R $} & \multicolumn{1}{c}{$\rm \rho_{c}$} & \multicolumn{1}{c}{$\rm \beta$} \\
			\hline
            \multirow[t]{3}{*}{BG-C1} & No cut-off & 0.001 & 1.776 & 9.012 & 58.881 & 7.953 \\
            & $k_{\rm min}=\pi /R$ & 0.001 & 0.068 & 1.762 & 9.083 & 1.724 \\
            & $k_{\rm min}=\pi /R_{\rm eff}$ & 0.001 & 1.140 & 7.158 & 28.926 & 7.744

		\end{tabular}
	\end{ruledtabular}
	\label{tab:table2}
\end{table}

\section{Concluding Remarks} \label{sec:remarks}
Compact objects composed of either fermionic or bosonic matter have been employed for studying the configurational entropy  as a means of stability. The aforementioned quantity should be in alignment with the stability criterion of TM method. It is important to note that the existence of a stable configuration is a property of gravity and independent of the EoS. However, the specific location of the stability point is influenced by the underlying EoS~\cite{Glendenning-2000b}. Considering the above points, one might expect, as suggested in Ref.~\cite{Gleiser-2015a}, that the minimization of the CE is a consequence of gravity within the framework of general relativity. If this is the case, the relevant minimum (which should be a total minimum) should be independent of the applied EoS. However, our findings indicate that this is not universally true for compact objects. 

The CE was studied in light of the gravitational mass, radius, central density and compactness. The TM and the CE method for the location of the stability point converge, with a good or moderate accuracy, for the three out of four quantities under consideration. In fact, the most accurate prediction lies with the maximum mass, where the difference reaches values lower than 5\%. In addition, a quantity with good accuracy is also the corresponding radius with errors up to 15\%, while in the majority of the cases, the error is lower than 10\%. As a result, the combination of the aforementioned macroscopic quantities, which is the compactness, is also a quantity with accuracy lying on values lower than 12\%. These three quantities are decisive indicators of the macroscopic quantities of compact stars and leading to the ultimate result that in a macroscopic scale, the two methods for locating the stability point, are in agreement.

The last quantity under consideration, which is the central density, leads to enormous amounts of error that can reach values up to 100\%. From this point of view, the two methods contradict each other rendering the central density unreliably calculated through the CE method. However, it is important to note that, while the previous analysis may result in significant errors in certain cases, the situation is different for boson stars. By carefully selecting an appropriate momentum cut-off, the convergence between the two methods can be greatly improved across all four quantities under consideration. This highlights the critical role of the momentum cut-off in mitigating errors and enhancing the overall accuracy of the results.

The above result does not agree with what was recently found that the stability by the minimization of the CE, concerning neutron stars and quark stars, does not have, at least quantitatively, universal validity. A possible explanation, at least for neutron stars, is that the existence of the crust, which has a special constitutive explanation, has a dramatic effect on locating the stability point by the CE minimization method. On the other hand, in the case of quark stars, where there is no crust, the failure of the method does not currently have a solid explanation. Thus, the accurate prediction of the stability point is not only related to the uniformity of the EoS, such as in the case of interacting Fermi and boson gas, where no crust is added, but also to its specific form.

In conclusion, the CE method can be used as a qualitative, rather than a quantitative, tool to macroscopically locate the instability region of certain configurations of compact objects. Finally, although the CE method is an alternative approach for exploring the instability regions of compact objects, the dependence on the specific EoS and the internal structure of the compact star are factors with a decisive role in the validity of the approach.

\section*{Acknowledgments}
The authors would like to thank Dr. Nan Jiang for correspondence and useful comments. All numerical calculations were performed on a workstation equipped with 2 Intel Xeon Gold 6140 Processors (72 cpu cores in total) provided by the MSc program “Computational Physics” of the Physics Department, Aristotle University of Thessaloniki. This work was supported by the Croatian Science Foundation under Project No. HRZZ-MOBDOL-12-2023-6026, by the Croatian Science Foundation under Project No. IP-2022-10-7773, and by the Czech Science Foundation (GACR Contract No. 21-24281S). 

\bibliography{koliogiannis_bib}

\begin{thebibliography}{59}%
\makeatletter
\providecommand \@ifxundefined [1]{%
 \@ifx{#1\undefined}
}%
\providecommand \@ifnum [1]{%
 \ifnum #1\expandafter \@firstoftwo
 \else \expandafter \@secondoftwo
 \fi
}%
\providecommand \@ifx [1]{%
 \ifx #1\expandafter \@firstoftwo
 \else \expandafter \@secondoftwo
 \fi
}%
\providecommand \natexlab [1]{#1}%
\providecommand \enquote  [1]{``#1''}%
\providecommand \bibnamefont  [1]{#1}%
\providecommand \bibfnamefont [1]{#1}%
\providecommand \citenamefont [1]{#1}%
\providecommand \href@noop [0]{\@secondoftwo}%
\providecommand \href [0]{\begingroup \@sanitize@url \@href}%
\providecommand \@href[1]{\@@startlink{#1}\@@href}%
\providecommand \@@href[1]{\endgroup#1\@@endlink}%
\providecommand \@sanitize@url [0]{\catcode `\\12\catcode `\$12\catcode
  `\&12\catcode `\#12\catcode `\^12\catcode `\_12\catcode `\%12\relax}%
\providecommand \@@startlink[1]{}%
\providecommand \@@endlink[0]{}%
\providecommand \url  [0]{\begingroup\@sanitize@url \@url }%
\providecommand \@url [1]{\endgroup\@href {#1}{\urlprefix }}%
\providecommand \urlprefix  [0]{URL }%
\providecommand \Eprint [0]{\href }%
\providecommand \doibase [0]{https://doi.org/}%
\providecommand \selectlanguage [0]{\@gobble}%
\providecommand \bibinfo  [0]{\@secondoftwo}%
\providecommand \bibfield  [0]{\@secondoftwo}%
\providecommand \translation [1]{[#1]}%
\providecommand \BibitemOpen [0]{}%
\providecommand \bibitemStop [0]{}%
\providecommand \bibitemNoStop [0]{.\EOS\space}%
\providecommand \EOS [0]{\spacefactor3000\relax}%
\providecommand \BibitemShut  [1]{\csname bibitem#1\endcsname}%
\let\auto@bib@innerbib\@empty
\bibitem [{\citenamefont {Sañudo}\ and\ \citenamefont
  {Pacheco}(2009)}]{Sanudo-2009}%
  \BibitemOpen
  \bibfield  {author} {\bibinfo {author} {\bibfnamefont {J.}~\bibnamefont
  {Sañudo}}\ and\ \bibinfo {author} {\bibfnamefont {A.}~\bibnamefont
  {Pacheco}},\ }\bibfield  {title} {\bibinfo {title} {Complexity and
  white-dwarf structure},\ }\href
  {https://doi.org/https://doi.org/10.1016/j.physleta.2009.01.008} {\bibfield
  {journal} {\bibinfo  {journal} {Phys. Lett. A}\ }\textbf {\bibinfo {volume}
  {373}},\ \bibinfo {pages} {807} (\bibinfo {year} {2009})}\BibitemShut
  {NoStop}%
\bibitem [{\citenamefont {Chatzisavvas}\ \emph {et~al.}(2009)\citenamefont
  {Chatzisavvas}, \citenamefont {Psonis}, \citenamefont {Panos},\ and\
  \citenamefont {Moustakidis}}]{Moustakidis-2009}%
  \BibitemOpen
  \bibfield  {author} {\bibinfo {author} {\bibfnamefont {K.}~\bibnamefont
  {Chatzisavvas}}, \bibinfo {author} {\bibfnamefont {V.}~\bibnamefont
  {Psonis}}, \bibinfo {author} {\bibfnamefont {C.}~\bibnamefont {Panos}},\ and\
  \bibinfo {author} {\bibfnamefont {C.}~\bibnamefont {Moustakidis}},\
  }\bibfield  {title} {\bibinfo {title} {Complexity and neutron star
  structure},\ }\href
  {https://doi.org/https://doi.org/10.1016/j.physleta.2009.08.042} {\bibfield
  {journal} {\bibinfo  {journal} {Physics Letters A}\ }\textbf {\bibinfo
  {volume} {373}},\ \bibinfo {pages} {3901} (\bibinfo {year}
  {2009})}\BibitemShut {NoStop}%
\bibitem [{\citenamefont {{de Avellar}}\ and\ \citenamefont
  {Horvath}(2012)}]{de_Avellar-2012}%
  \BibitemOpen
  \bibfield  {author} {\bibinfo {author} {\bibfnamefont {M.}~\bibnamefont {{de
  Avellar}}}\ and\ \bibinfo {author} {\bibfnamefont {J.}~\bibnamefont
  {Horvath}},\ }\bibfield  {title} {\bibinfo {title} {Entropy, complexity and
  disequilibrium in compact stars},\ }\href
  {https://doi.org/https://doi.org/10.1016/j.physleta.2012.02.012} {\bibfield
  {journal} {\bibinfo  {journal} {Phys. Lett. A}\ }\textbf {\bibinfo {volume}
  {376}},\ \bibinfo {pages} {1085} (\bibinfo {year} {2012})}\BibitemShut
  {NoStop}%
\bibitem [{\citenamefont {{de Avellar}}\ \emph {et~al.}(2014)\citenamefont {{de
  Avellar}}, \citenamefont {{de Souza}}, \citenamefont {Horvath},\ and\
  \citenamefont {Paret}}]{de_Avellar-2014}%
  \BibitemOpen
  \bibfield  {author} {\bibinfo {author} {\bibfnamefont {M.}~\bibnamefont {{de
  Avellar}}}, \bibinfo {author} {\bibfnamefont {R.}~\bibnamefont {{de Souza}}},
  \bibinfo {author} {\bibfnamefont {J.}~\bibnamefont {Horvath}},\ and\ \bibinfo
  {author} {\bibfnamefont {D.}~\bibnamefont {Paret}},\ }\bibfield  {title}
  {\bibinfo {title} {Information theoretical methods as discerning quantifiers
  of the equations of state of neutron stars},\ }\href
  {https://doi.org/https://doi.org/10.1016/j.physleta.2014.10.011} {\bibfield
  {journal} {\bibinfo  {journal} {Phys. Lett. A}\ }\textbf {\bibinfo {volume}
  {378}},\ \bibinfo {pages} {3481} (\bibinfo {year} {2014})}\BibitemShut
  {NoStop}%
\bibitem [{\citenamefont {Adhitya}\ and\ \citenamefont
  {Sulaksono}(2020)}]{Adhitya_2020}%
  \BibitemOpen
  \bibfield  {author} {\bibinfo {author} {\bibfnamefont {H.}~\bibnamefont
  {Adhitya}}\ and\ \bibinfo {author} {\bibfnamefont {A.}~\bibnamefont
  {Sulaksono}},\ }\bibfield  {title} {\bibinfo {title} {Complexity and neutron
  stars with crust and hyperon core},\ }\href
  {https://doi.org/10.1088/1742-6596/1572/1/012012} {\bibfield  {journal}
  {\bibinfo  {journal} {Journal of Physics: Conference Series}\ }\textbf
  {\bibinfo {volume} {1572}},\ \bibinfo {pages} {012012} (\bibinfo {year}
  {2020})}\BibitemShut {NoStop}%
\bibitem [{\citenamefont {Contreras}\ and\ \citenamefont
  {Fuenmayor}(2021)}]{Contreras-2021}%
  \BibitemOpen
  \bibfield  {author} {\bibinfo {author} {\bibfnamefont {E.}~\bibnamefont
  {Contreras}}\ and\ \bibinfo {author} {\bibfnamefont {E.}~\bibnamefont
  {Fuenmayor}},\ }\bibfield  {title} {\bibinfo {title} {Gravitational cracking
  and complexity in the framework of gravitational decoupling},\ }\href
  {https://doi.org/10.1103/PhysRevD.103.124065} {\bibfield  {journal} {\bibinfo
   {journal} {Phys. Rev. D}\ }\textbf {\bibinfo {volume} {103}},\ \bibinfo
  {pages} {124065} (\bibinfo {year} {2021})}\BibitemShut {NoStop}%
\bibitem [{\citenamefont {Posada}\ \emph {et~al.}(2021)\citenamefont {Posada},
  \citenamefont {Hlad\'{\i}k},\ and\ \citenamefont
  {Stuchl\'{\i}k}}]{Posada-2021}%
  \BibitemOpen
  \bibfield  {author} {\bibinfo {author} {\bibfnamefont {C.}~\bibnamefont
  {Posada}}, \bibinfo {author} {\bibfnamefont {J.}~\bibnamefont
  {Hlad\'{\i}k}},\ and\ \bibinfo {author} {\bibfnamefont {Z.~c.~v.}\
  \bibnamefont {Stuchl\'{\i}k}},\ }\bibfield  {title} {\bibinfo {title}
  {Dynamical stability of the modified {T}olman {VII} solution},\ }\href
  {https://doi.org/10.1103/PhysRevD.103.104067} {\bibfield  {journal} {\bibinfo
   {journal} {Phys. Rev. D}\ }\textbf {\bibinfo {volume} {103}},\ \bibinfo
  {pages} {104067} (\bibinfo {year} {2021})}\BibitemShut {NoStop}%
\bibitem [{\citenamefont {Herrera}(2018)}]{Herrera-2018a}%
  \BibitemOpen
  \bibfield  {author} {\bibinfo {author} {\bibfnamefont {L.}~\bibnamefont
  {Herrera}},\ }\bibfield  {title} {\bibinfo {title} {New definition of
  complexity for self-gravitating fluid distributions: {T}he spherically
  symmetric, static case},\ }\href {https://doi.org/10.1103/PhysRevD.97.044010}
  {\bibfield  {journal} {\bibinfo  {journal} {Phys. Rev. D}\ }\textbf {\bibinfo
  {volume} {97}},\ \bibinfo {pages} {044010} (\bibinfo {year}
  {2018})}\BibitemShut {NoStop}%
\bibitem [{\citenamefont {Herrera}\ \emph {et~al.}(2018)\citenamefont
  {Herrera}, \citenamefont {Di~Prisco},\ and\ \citenamefont
  {Ospino}}]{Herrera-2018b}%
  \BibitemOpen
  \bibfield  {author} {\bibinfo {author} {\bibfnamefont {L.}~\bibnamefont
  {Herrera}}, \bibinfo {author} {\bibfnamefont {A.}~\bibnamefont {Di~Prisco}},\
  and\ \bibinfo {author} {\bibfnamefont {J.}~\bibnamefont {Ospino}},\
  }\bibfield  {title} {\bibinfo {title} {Definition of complexity for dynamical
  spherically symmetric dissipative self-gravitating fluid distributions},\
  }\href {https://doi.org/10.1103/PhysRevD.98.104059} {\bibfield  {journal}
  {\bibinfo  {journal} {Phys. Rev. D}\ }\textbf {\bibinfo {volume} {98}},\
  \bibinfo {pages} {104059} (\bibinfo {year} {2018})}\BibitemShut {NoStop}%
\bibitem [{\citenamefont {Herrera}\ \emph
  {et~al.}(2019{\natexlab{a}})\citenamefont {Herrera}, \citenamefont
  {Di~Prisco},\ and\ \citenamefont {Ospino}}]{Herrera-2019a}%
  \BibitemOpen
  \bibfield  {author} {\bibinfo {author} {\bibfnamefont {L.}~\bibnamefont
  {Herrera}}, \bibinfo {author} {\bibfnamefont {A.}~\bibnamefont {Di~Prisco}},\
  and\ \bibinfo {author} {\bibfnamefont {J.}~\bibnamefont {Ospino}},\
  }\bibfield  {title} {\bibinfo {title} {Complexity factors for axially
  symmetric static sources},\ }\href
  {https://doi.org/10.1103/PhysRevD.99.044049} {\bibfield  {journal} {\bibinfo
  {journal} {Phys. Rev. D}\ }\textbf {\bibinfo {volume} {99}},\ \bibinfo
  {pages} {044049} (\bibinfo {year} {2019}{\natexlab{a}})}\BibitemShut
  {NoStop}%
\bibitem [{\citenamefont {Herrera}\ \emph
  {et~al.}(2019{\natexlab{b}})\citenamefont {Herrera}, \citenamefont
  {Di~Prisco},\ and\ \citenamefont {Carot}}]{Herrera-2019b}%
  \BibitemOpen
  \bibfield  {author} {\bibinfo {author} {\bibfnamefont {L.}~\bibnamefont
  {Herrera}}, \bibinfo {author} {\bibfnamefont {A.}~\bibnamefont {Di~Prisco}},\
  and\ \bibinfo {author} {\bibfnamefont {J.}~\bibnamefont {Carot}},\ }\bibfield
   {title} {\bibinfo {title} {Complexity of the {B}ondi {M}etric},\ }\href
  {https://doi.org/10.1103/PhysRevD.99.124028} {\bibfield  {journal} {\bibinfo
  {journal} {Phys. Rev. D}\ }\textbf {\bibinfo {volume} {99}},\ \bibinfo
  {pages} {124028} (\bibinfo {year} {2019}{\natexlab{b}})}\BibitemShut
  {NoStop}%
\bibitem [{\citenamefont {Sharif}\ and\ \citenamefont
  {Butt}(2018{\natexlab{a}})}]{Sharif-2018a}%
  \BibitemOpen
  \bibfield  {author} {\bibinfo {author} {\bibfnamefont {M.}~\bibnamefont
  {Sharif}}\ and\ \bibinfo {author} {\bibfnamefont {I.}~\bibnamefont {Butt}},\
  }\bibfield  {title} {\bibinfo {title} {Complexity factor for charged
  spherical system},\ }\href {https://doi.org/10.1140/epjc/s10052-018-6121-5}
  {\bibfield  {journal} {\bibinfo  {journal} {Eur. Phys. J. C}\ }\textbf
  {\bibinfo {volume} {78}},\ \bibinfo {pages} {688} (\bibinfo {year}
  {2018}{\natexlab{a}})}\BibitemShut {NoStop}%
\bibitem [{\citenamefont {Sharif}\ and\ \citenamefont
  {Butt}(2018{\natexlab{b}})}]{Sharif-2018b}%
  \BibitemOpen
  \bibfield  {author} {\bibinfo {author} {\bibfnamefont {M.}~\bibnamefont
  {Sharif}}\ and\ \bibinfo {author} {\bibfnamefont {I.~I.}\ \bibnamefont
  {Butt}},\ }\bibfield  {title} {\bibinfo {title} {Complexity factor for static
  cylindrical system},\ }\href {https://doi.org/10.1140/epjc/s10052-018-6330-y}
  {\bibfield  {journal} {\bibinfo  {journal} {Eur. Phys. J. C}\ }\textbf
  {\bibinfo {volume} {78}},\ \bibinfo {pages} {850} (\bibinfo {year}
  {2018}{\natexlab{b}})}\BibitemShut {NoStop}%
\bibitem [{\citenamefont {Sharif}\ \emph {et~al.}(2019)\citenamefont {Sharif},
  \citenamefont {Majid},\ and\ \citenamefont {Nasir}}]{Sharif-2019}%
  \BibitemOpen
  \bibfield  {author} {\bibinfo {author} {\bibfnamefont {M.}~\bibnamefont
  {Sharif}}, \bibinfo {author} {\bibfnamefont {A.}~\bibnamefont {Majid}},\ and\
  \bibinfo {author} {\bibfnamefont {M.~M.~M.}\ \bibnamefont {Nasir}},\
  }\bibfield  {title} {\bibinfo {title} {Complexity factor for self-gravitating
  system in modified {G}auss–{B}onnet gravity},\ }\href
  {https://doi.org/10.1142/S0217751X19502105} {\bibfield  {journal} {\bibinfo
  {journal} {Int. J. Mod. Phys. A}\ }\textbf {\bibinfo {volume} {34}},\
  \bibinfo {pages} {1950210} (\bibinfo {year} {2019})}\BibitemShut {NoStop}%
\bibitem [{\citenamefont {Sharif}\ and\ \citenamefont
  {Hassan}(2022)}]{Sharif-2022}%
  \BibitemOpen
  \bibfield  {author} {\bibinfo {author} {\bibfnamefont {M.}~\bibnamefont
  {Sharif}}\ and\ \bibinfo {author} {\bibfnamefont {K.}~\bibnamefont
  {Hassan}},\ }\bibfield  {title} {\bibinfo {title} {Complexity of dynamical
  cylindrical system in f({G, T}) gravity},\ }\href
  {https://doi.org/10.1142/S0217732322500274} {\bibfield  {journal} {\bibinfo
  {journal} {Mod. Phys. Lett. A}\ }\textbf {\bibinfo {volume} {37}},\ \bibinfo
  {pages} {2250027} (\bibinfo {year} {2022})}\BibitemShut {NoStop}%
\bibitem [{\citenamefont {Yousaf}\ \emph
  {et~al.}(2020{\natexlab{a}})\citenamefont {Yousaf}, \citenamefont {Bhatti},\
  and\ \citenamefont {Naseer}}]{Yousaf-2020a}%
  \BibitemOpen
  \bibfield  {author} {\bibinfo {author} {\bibfnamefont {Z.}~\bibnamefont
  {Yousaf}}, \bibinfo {author} {\bibfnamefont {M.~Z.}\ \bibnamefont {Bhatti}},\
  and\ \bibinfo {author} {\bibfnamefont {T.}~\bibnamefont {Naseer}},\
  }\bibfield  {title} {\bibinfo {title} {Study of static charged spherical
  structure in f({R,T,Q}) gravity},\ }\href
  {https://doi.org/10.1140/epjp/s13360-020-00332-9} {\bibfield  {journal}
  {\bibinfo  {journal} {Eur. Phys. J. Plus}\ }\textbf {\bibinfo {volume}
  {135}},\ \bibinfo {pages} {323} (\bibinfo {year}
  {2020}{\natexlab{a}})}\BibitemShut {NoStop}%
\bibitem [{\citenamefont {Yousaf}\ \emph
  {et~al.}(2020{\natexlab{b}})\citenamefont {Yousaf}, \citenamefont {Khlopov},
  \citenamefont {Bhatti},\ and\ \citenamefont {Naseer}}]{Yousaf-2020b}%
  \BibitemOpen
  \bibfield  {author} {\bibinfo {author} {\bibfnamefont {Z.}~\bibnamefont
  {Yousaf}}, \bibinfo {author} {\bibfnamefont {M.~Y.}\ \bibnamefont {Khlopov}},
  \bibinfo {author} {\bibfnamefont {M.~Z.}\ \bibnamefont {Bhatti}},\ and\
  \bibinfo {author} {\bibfnamefont {T.}~\bibnamefont {Naseer}},\ }\bibfield
  {title} {\bibinfo {title} {{Influence of modification of gravity on the
  complexity factor of static spherical structures}},\ }\href
  {https://doi.org/10.1093/mnras/staa1470} {\bibfield  {journal} {\bibinfo
  {journal} {Mon. Not. R. Astron. Soc.}\ }\textbf {\bibinfo {volume} {495}},\
  \bibinfo {pages} {4334} (\bibinfo {year} {2020}{\natexlab{b}})}\BibitemShut
  {NoStop}%
\bibitem [{\citenamefont {Yousaf}\ \emph
  {et~al.}(2020{\natexlab{c}})\citenamefont {Yousaf}, \citenamefont {Bhatti},\
  and\ \citenamefont {Naseer}}]{Yousaf-2020c}%
  \BibitemOpen
  \bibfield  {author} {\bibinfo {author} {\bibfnamefont {Z.}~\bibnamefont
  {Yousaf}}, \bibinfo {author} {\bibfnamefont {M.~Z.}\ \bibnamefont {Bhatti}},\
  and\ \bibinfo {author} {\bibfnamefont {T.}~\bibnamefont {Naseer}},\
  }\bibfield  {title} {\bibinfo {title} {Measure of complexity for dynamical
  self-gravitating structures},\ }\href
  {https://doi.org/10.1142/S0218271820500613} {\bibfield  {journal} {\bibinfo
  {journal} {Int. J. Mod. Phys. D}\ }\textbf {\bibinfo {volume} {29}},\
  \bibinfo {pages} {2050061} (\bibinfo {year}
  {2020}{\natexlab{c}})}\BibitemShut {NoStop}%
\bibitem [{\citenamefont {Yousaf}\ \emph {et~al.}(2021)\citenamefont {Yousaf},
  \citenamefont {Bamba}, \citenamefont {Bhatti},\ and\ \citenamefont
  {Hassan}}]{Yousaf-2021}%
  \BibitemOpen
  \bibfield  {author} {\bibinfo {author} {\bibfnamefont {Z.}~\bibnamefont
  {Yousaf}}, \bibinfo {author} {\bibfnamefont {K.}~\bibnamefont {Bamba}},
  \bibinfo {author} {\bibfnamefont {M.}~\bibnamefont {Bhatti}},\ and\ \bibinfo
  {author} {\bibfnamefont {K.}~\bibnamefont {Hassan}},\ }\bibfield  {title}
  {\bibinfo {title} {Measure of complexity in self-gravitating systems using
  structure scalars},\ }\href
  {https://doi.org/https://doi.org/10.1016/j.newast.2020.101541} {\bibfield
  {journal} {\bibinfo  {journal} {New Astronomy}\ }\textbf {\bibinfo {volume}
  {84}},\ \bibinfo {pages} {101541} (\bibinfo {year} {2021})}\BibitemShut
  {NoStop}%
\bibitem [{\citenamefont {Yousaf}\ \emph {et~al.}(2022)\citenamefont {Yousaf},
  \citenamefont {Bhatti},\ and\ \citenamefont {Nasir}}]{Yousaf-2022}%
  \BibitemOpen
  \bibfield  {author} {\bibinfo {author} {\bibfnamefont {Z.}~\bibnamefont
  {Yousaf}}, \bibinfo {author} {\bibfnamefont {M.}~\bibnamefont {Bhatti}},\
  and\ \bibinfo {author} {\bibfnamefont {M.}~\bibnamefont {Nasir}},\ }\bibfield
   {title} {\bibinfo {title} {On the study of complexity for charged
  self-gravitating systems},\ }\href
  {https://doi.org/https://doi.org/10.1016/j.cjph.2022.01.005} {\bibfield
  {journal} {\bibinfo  {journal} {Chin. J. Phys.}\ }\textbf {\bibinfo {volume}
  {77}},\ \bibinfo {pages} {2078} (\bibinfo {year} {2022})}\BibitemShut
  {NoStop}%
\bibitem [{\citenamefont {Gleiser}\ and\ \citenamefont
  {Stamatopoulos}(2012)}]{Gleiser-2012}%
  \BibitemOpen
  \bibfield  {author} {\bibinfo {author} {\bibfnamefont {M.}~\bibnamefont
  {Gleiser}}\ and\ \bibinfo {author} {\bibfnamefont {N.}~\bibnamefont
  {Stamatopoulos}},\ }\bibfield  {title} {\bibinfo {title} {Entropic measure
  for localized energy configurations: {K}inks, bounces, and bubbles},\ }\href
  {https://doi.org/https://doi.org/10.1016/j.physletb.2012.05.064} {\bibfield
  {journal} {\bibinfo  {journal} {Phys. Lett. B}\ }\textbf {\bibinfo {volume}
  {713}},\ \bibinfo {pages} {304} (\bibinfo {year} {2012})}\BibitemShut
  {NoStop}%
\bibitem [{\citenamefont {Gleiser}\ and\ \citenamefont
  {Sowinski}(2013)}]{Gleiser-2013}%
  \BibitemOpen
  \bibfield  {author} {\bibinfo {author} {\bibfnamefont {M.}~\bibnamefont
  {Gleiser}}\ and\ \bibinfo {author} {\bibfnamefont {D.}~\bibnamefont
  {Sowinski}},\ }\bibfield  {title} {\bibinfo {title} {Information-entropic
  stability bound for compact objects: {A}pplication to {Q}-balls and the
  {C}handrasekhar limit of polytropes},\ }\href
  {https://doi.org/https://doi.org/10.1016/j.physletb.2013.10.005} {\bibfield
  {journal} {\bibinfo  {journal} {Phys. Lett. B}\ }\textbf {\bibinfo {volume}
  {727}},\ \bibinfo {pages} {272} (\bibinfo {year} {2013})}\BibitemShut
  {NoStop}%
\bibitem [{\citenamefont {Gleiser}\ and\ \citenamefont
  {Jiang}(2015)}]{Gleiser-2015a}%
  \BibitemOpen
  \bibfield  {author} {\bibinfo {author} {\bibfnamefont {M.}~\bibnamefont
  {Gleiser}}\ and\ \bibinfo {author} {\bibfnamefont {N.}~\bibnamefont
  {Jiang}},\ }\bibfield  {title} {\bibinfo {title} {Stability bounds on compact
  astrophysical objects from information-entropic measure},\ }\href
  {https://doi.org/10.1103/PhysRevD.92.044046} {\bibfield  {journal} {\bibinfo
  {journal} {Phys. Rev. D}\ }\textbf {\bibinfo {volume} {92}},\ \bibinfo
  {pages} {044046} (\bibinfo {year} {2015})}\BibitemShut {NoStop}%
\bibitem [{\citenamefont {Koliogiannis}\ \emph {et~al.}(2023)\citenamefont
  {Koliogiannis}, \citenamefont {Tsalis}, \citenamefont {Panos},\ and\
  \citenamefont {Moustakidis}}]{Koliogiannis-2023}%
  \BibitemOpen
  \bibfield  {author} {\bibinfo {author} {\bibfnamefont {P.~S.}\ \bibnamefont
  {Koliogiannis}}, \bibinfo {author} {\bibfnamefont {G.~A.}\ \bibnamefont
  {Tsalis}}, \bibinfo {author} {\bibfnamefont {C.~P.}\ \bibnamefont {Panos}},\
  and\ \bibinfo {author} {\bibfnamefont {C.~C.}\ \bibnamefont {Moustakidis}},\
  }\bibfield  {title} {\bibinfo {title} {Configurational entropy as a probe of
  the stability condition of compact objects},\ }\href
  {https://doi.org/10.1103/PhysRevD.107.044069} {\bibfield  {journal} {\bibinfo
   {journal} {Phys. Rev. D}\ }\textbf {\bibinfo {volume} {107}},\ \bibinfo
  {pages} {044069} (\bibinfo {year} {2023})}\BibitemShut {NoStop}%
\bibitem [{\citenamefont {Gleiser}\ and\ \citenamefont
  {Sowinski}(2015)}]{Gleiser-2015b}%
  \BibitemOpen
  \bibfield  {author} {\bibinfo {author} {\bibfnamefont {M.}~\bibnamefont
  {Gleiser}}\ and\ \bibinfo {author} {\bibfnamefont {D.}~\bibnamefont
  {Sowinski}},\ }\bibfield  {title} {\bibinfo {title} {Information-entropic
  signature of the critical point},\ }\href
  {https://doi.org/https://doi.org/10.1016/j.physletb.2015.05.058} {\bibfield
  {journal} {\bibinfo  {journal} {Phys. Lett. B}\ }\textbf {\bibinfo {volume}
  {747}},\ \bibinfo {pages} {125} (\bibinfo {year} {2015})}\BibitemShut
  {NoStop}%
\bibitem [{\citenamefont {Braga}(2019)}]{Braga-2019}%
  \BibitemOpen
  \bibfield  {author} {\bibinfo {author} {\bibfnamefont {N.~R.}\ \bibnamefont
  {Braga}},\ }\bibfield  {title} {\bibinfo {title} {Information versus
  stability in an anti-de {S}itter black hole},\ }\href
  {https://doi.org/https://doi.org/10.1016/j.physletb.2019.134919} {\bibfield
  {journal} {\bibinfo  {journal} {Phys. Lett. B}\ }\textbf {\bibinfo {volume}
  {797}},\ \bibinfo {pages} {134919} (\bibinfo {year} {2019})}\BibitemShut
  {NoStop}%
\bibitem [{\citenamefont {Braga}\ and\ \citenamefont
  {da~Mata}(2020)}]{Braga-2020}%
  \BibitemOpen
  \bibfield  {author} {\bibinfo {author} {\bibfnamefont {N.~R.~F.}\
  \bibnamefont {Braga}}\ and\ \bibinfo {author} {\bibfnamefont
  {R.}~\bibnamefont {da~Mata}},\ }\bibfield  {title} {\bibinfo {title}
  {Configuration entropy for quarkonium in a finite density plasma},\ }\href
  {https://doi.org/10.1103/PhysRevD.101.105016} {\bibfield  {journal} {\bibinfo
   {journal} {Phys. Rev. D}\ }\textbf {\bibinfo {volume} {101}},\ \bibinfo
  {pages} {105016} (\bibinfo {year} {2020})}\BibitemShut {NoStop}%
\bibitem [{\citenamefont {Alexander}\ \emph {et~al.}(2018)\citenamefont
  {Alexander}, \citenamefont {Yagi},\ and\ \citenamefont {Yunes}}]{Yunes-2018}%
  \BibitemOpen
  \bibfield  {author} {\bibinfo {author} {\bibfnamefont {S.~H.}\ \bibnamefont
  {Alexander}}, \bibinfo {author} {\bibfnamefont {K.}~\bibnamefont {Yagi}},\
  and\ \bibinfo {author} {\bibfnamefont {N.}~\bibnamefont {Yunes}},\ }\bibfield
   {title} {\bibinfo {title} {An entropy-area law for neutron stars near the
  black hole threshold},\ }\href {https://doi.org/10.1088/1361-6382/aaf14b}
  {\bibfield  {journal} {\bibinfo  {journal} {Class. Quantum Grav.}\ }\textbf
  {\bibinfo {volume} {36}},\ \bibinfo {pages} {015010} (\bibinfo {year}
  {2018})}\BibitemShut {NoStop}%
\bibitem [{\citenamefont {{da Rocha}}(2021)}]{Rocha-2021}%
  \BibitemOpen
  \bibfield  {author} {\bibinfo {author} {\bibfnamefont {R.}~\bibnamefont {{da
  Rocha}}},\ }\bibfield  {title} {\bibinfo {title} {Ad{S} graviton stars and
  differential configurational entropy},\ }\href
  {https://doi.org/https://doi.org/10.1016/j.physletb.2021.136729} {\bibfield
  {journal} {\bibinfo  {journal} {Phys. Lett. B}\ }\textbf {\bibinfo {volume}
  {823}},\ \bibinfo {pages} {136729} (\bibinfo {year} {2021})}\BibitemShut
  {NoStop}%
\bibitem [{\citenamefont {Karapetyan}(2018)}]{Karapetyan-2018}%
  \BibitemOpen
  \bibfield  {author} {\bibinfo {author} {\bibfnamefont {G.}~\bibnamefont
  {Karapetyan}},\ }\bibfield  {title} {\bibinfo {title} {The nuclear
  configurational entropy approach to dynamical {QCD} effects},\ }\href
  {https://doi.org/https://doi.org/10.1016/j.physletb.2018.09.058} {\bibfield
  {journal} {\bibinfo  {journal} {Phys. Lett. B}\ }\textbf {\bibinfo {volume}
  {786}},\ \bibinfo {pages} {418} (\bibinfo {year} {2018})}\BibitemShut
  {NoStop}%
\bibitem [{\citenamefont {Correa}\ \emph {et~al.}(2016)\citenamefont {Correa},
  \citenamefont {Moraes}, \citenamefont {de~Souza~Dutra}, \citenamefont
  {de~Paula},\ and\ \citenamefont {Frederico}}]{Correa-2016}%
  \BibitemOpen
  \bibfield  {author} {\bibinfo {author} {\bibfnamefont {R.~A.~C.}\
  \bibnamefont {Correa}}, \bibinfo {author} {\bibfnamefont {P.~H. R.~S.}\
  \bibnamefont {Moraes}}, \bibinfo {author} {\bibfnamefont {A.}~\bibnamefont
  {de~Souza~Dutra}}, \bibinfo {author} {\bibfnamefont {W.}~\bibnamefont
  {de~Paula}},\ and\ \bibinfo {author} {\bibfnamefont {T.}~\bibnamefont
  {Frederico}},\ }\bibfield  {title} {\bibinfo {title} {Configurational entropy
  as a constraint for {G}auss-{B}onnet braneworld models},\ }\href
  {https://doi.org/10.1103/PhysRevD.94.083509} {\bibfield  {journal} {\bibinfo
  {journal} {Phys. Rev. D}\ }\textbf {\bibinfo {volume} {94}},\ \bibinfo
  {pages} {083509} (\bibinfo {year} {2016})}\BibitemShut {NoStop}%
\bibitem [{\citenamefont {Barreto}\ \emph {et~al.}(2022)\citenamefont
  {Barreto}, \citenamefont {Herrera–Aguilar},\ and\ \citenamefont {{da
  Rocha}}}]{Baretto-2022}%
  \BibitemOpen
  \bibfield  {author} {\bibinfo {author} {\bibfnamefont {W.}~\bibnamefont
  {Barreto}}, \bibinfo {author} {\bibfnamefont {A.}~\bibnamefont
  {Herrera–Aguilar}},\ and\ \bibinfo {author} {\bibfnamefont
  {R.}~\bibnamefont {{da Rocha}}},\ }\bibfield  {title} {\bibinfo {title}
  {Configurational entropy of generalized sine–{G}ordon-type models},\ }\href
  {https://doi.org/https://doi.org/10.1016/j.aop.2022.169142} {\bibfield
  {journal} {\bibinfo  {journal} {Ann. Phys.}\ }\textbf {\bibinfo {volume}
  {447}},\ \bibinfo {pages} {169142} (\bibinfo {year} {2022})}\BibitemShut
  {NoStop}%
\bibitem [{\citenamefont {Shapiro}\ and\ \citenamefont
  {Teukolsky}(1983)}]{Shapiro-1983}%
  \BibitemOpen
  \bibfield  {author} {\bibinfo {author} {\bibfnamefont {S.~L.}\ \bibnamefont
  {Shapiro}}\ and\ \bibinfo {author} {\bibfnamefont {S.~A.}\ \bibnamefont
  {Teukolsky}},\ }\href@noop {} {\emph {\bibinfo {title} {Black {H}oles,
  {W}hite {D}warfs, and {N}eutron {S}tars}}}\ (\bibinfo  {publisher} {John
  Wiley \& Sons},\ \bibinfo {address} {New York},\ \bibinfo {year}
  {1983})\BibitemShut {NoStop}%
\bibitem [{\citenamefont {Glendenning}(2000)}]{Glendenning-2000}%
  \BibitemOpen
  \bibfield  {author} {\bibinfo {author} {\bibfnamefont {N.}~\bibnamefont
  {Glendenning}},\ }\href@noop {} {\emph {\bibinfo {title} {Compact {S}tars:
  {N}uclear {P}hysics, {P}article {P}hysics, and {G}eneral {R}elativity}}}\
  (\bibinfo  {publisher} {Springer},\ \bibinfo {address} {Berlin},\ \bibinfo
  {year} {2000})\BibitemShut {NoStop}%
\bibitem [{\citenamefont {Haensel}\ \emph {et~al.}(2007)\citenamefont
  {Haensel}, \citenamefont {Potekhin},\ and\ \citenamefont
  {Yakovlev}}]{Haensel-2007}%
  \BibitemOpen
  \bibfield  {author} {\bibinfo {author} {\bibfnamefont {P.}~\bibnamefont
  {Haensel}}, \bibinfo {author} {\bibfnamefont {A.}~\bibnamefont {Potekhin}},\
  and\ \bibinfo {author} {\bibfnamefont {D.}~\bibnamefont {Yakovlev}},\
  }\href@noop {} {\emph {\bibinfo {title} {Neutron {S}tars 1: {E}quation of
  {S}tate and {S}tructure}}}\ (\bibinfo  {publisher} {Springer-Verlag},\
  \bibinfo {address} {New York},\ \bibinfo {year} {2007})\BibitemShut {NoStop}%
\bibitem [{\citenamefont {Zeldovich}\ and\ \citenamefont
  {Novikov}(1971)}]{Zeldovich-71}%
  \BibitemOpen
  \bibfield  {author} {\bibinfo {author} {\bibfnamefont {Y.}~\bibnamefont
  {Zeldovich}}\ and\ \bibinfo {author} {\bibfnamefont {I.}~\bibnamefont
  {Novikov}},\ }\href@noop {} {\emph {\bibinfo {title} {Stars and
  {R}elativity}}}\ (\bibinfo  {publisher} {Dover Publications},\ \bibinfo
  {address} {INC, Mineapolis New York},\ \bibinfo {year} {1971})\BibitemShut
  {NoStop}%
\bibitem [{\citenamefont {Weinberg}(1972)}]{Weinberg-72}%
  \BibitemOpen
  \bibfield  {author} {\bibinfo {author} {\bibfnamefont {S.}~\bibnamefont
  {Weinberg}},\ }\href@noop {} {\emph {\bibinfo {title} {Gravitation and
  {C}osmology: {P}rinciples and {A}pplications of the {G}eneral {T}heory of
  {R}elativity}}}\ (\bibinfo  {publisher} {Wiley},\ \bibinfo {address} {New
  York},\ \bibinfo {year} {1972})\BibitemShut {NoStop}%
\bibitem [{\citenamefont {Schutz}(1985)}]{Schutz-85}%
  \BibitemOpen
  \bibfield  {author} {\bibinfo {author} {\bibfnamefont {B.}~\bibnamefont
  {Schutz}},\ }\href@noop {} {\emph {\bibinfo {title} {A {F}irst {C}ourse in
  {G}eneral {R}elativity}}}\ (\bibinfo  {publisher} {Cambridge University
  Press},\ \bibinfo {address} {Cambridge},\ \bibinfo {year} {1985})\BibitemShut
  {NoStop}%
\bibitem [{\citenamefont {Schaffner-Bielich}(2020)}]{Bielich-2020}%
  \BibitemOpen
  \bibfield  {author} {\bibinfo {author} {\bibfnamefont {J.}~\bibnamefont
  {Schaffner-Bielich}},\ }\href@noop {} {\emph {\bibinfo {title} {Compact
  {S}tar {P}hysics}}}\ (\bibinfo  {publisher} {Cambridge University Press},\
  \bibinfo {address} {Cambridge},\ \bibinfo {year} {2020})\BibitemShut
  {NoStop}%
\bibitem [{\citenamefont
  {Chandrasekhar}(1964{\natexlab{a}})}]{Chandrasekhar-1964a}%
  \BibitemOpen
  \bibfield  {author} {\bibinfo {author} {\bibfnamefont {S.}~\bibnamefont
  {Chandrasekhar}},\ }\bibfield  {title} {\bibinfo {title} {The {D}ynamical
  {I}nstability of {G}aseous {M}asses {A}pproaching the {S}chwarzschild limit
  in {G}eneral {R}elativity.},\ }\href {https://doi.org/10.1086/147938}
  {\bibfield  {journal} {\bibinfo  {journal} {Astrophys. J.}\ }\textbf
  {\bibinfo {volume} {140}},\ \bibinfo {pages} {417} (\bibinfo {year}
  {1964}{\natexlab{a}})}\BibitemShut {NoStop}%
\bibitem [{\citenamefont
  {Chandrasekhar}(1964{\natexlab{b}})}]{Chandrasekhar-1964b}%
  \BibitemOpen
  \bibfield  {author} {\bibinfo {author} {\bibfnamefont {S.}~\bibnamefont
  {Chandrasekhar}},\ }\bibfield  {title} {\bibinfo {title} {Dynamical
  {I}nstability of {G}aseous {M}asses {A}pproaching the {S}chwarzschild {L}imit
  in {G}eneral {R}elativity},\ }\href
  {https://doi.org/10.1103/PhysRevLett.12.114} {\bibfield  {journal} {\bibinfo
  {journal} {Phys. Rev. Lett.}\ }\textbf {\bibinfo {volume} {12}},\ \bibinfo
  {pages} {114} (\bibinfo {year} {1964}{\natexlab{b}})}\BibitemShut {NoStop}%
\bibitem [{\citenamefont {Liddle}\ and\ \citenamefont
  {Madsen}(1992)}]{Liddle-1992}%
  \BibitemOpen
  \bibfield  {author} {\bibinfo {author} {\bibfnamefont {A.~R.}\ \bibnamefont
  {Liddle}}\ and\ \bibinfo {author} {\bibfnamefont {M.~S.}\ \bibnamefont
  {Madsen}},\ }\bibfield  {title} {\bibinfo {title} {The structure and
  formation of boson stars},\ }\href
  {https://doi.org/10.1142/S0218271892000057} {\bibfield  {journal} {\bibinfo
  {journal} {Int. J. Mod. Phys. D}\ }\textbf {\bibinfo {volume} {01}},\
  \bibinfo {pages} {101} (\bibinfo {year} {1992})}\BibitemShut {NoStop}%
\bibitem [{\citenamefont {Liebling}\ and\ \citenamefont
  {Palenzuela}(2017)}]{Liebling-2017}%
  \BibitemOpen
  \bibfield  {author} {\bibinfo {author} {\bibfnamefont {S.~L.}\ \bibnamefont
  {Liebling}}\ and\ \bibinfo {author} {\bibfnamefont {C.}~\bibnamefont
  {Palenzuela}},\ }\bibfield  {title} {\bibinfo {title} {Dynamical boson
  stars},\ }\href {https://doi.org/10.1007/s41114-017-0007-y} {\bibfield
  {journal} {\bibinfo  {journal} {Living Rev. Relativ.}\ }\textbf {\bibinfo
  {volume} {20}},\ \bibinfo {pages} {5} (\bibinfo {year} {2017})}\BibitemShut
  {NoStop}%
\bibitem [{\citenamefont {Schunck}\ and\ \citenamefont
  {Mielke}(2003)}]{Schunck-2003}%
  \BibitemOpen
  \bibfield  {author} {\bibinfo {author} {\bibfnamefont {F.~E.}\ \bibnamefont
  {Schunck}}\ and\ \bibinfo {author} {\bibfnamefont {E.~W.}\ \bibnamefont
  {Mielke}},\ }\bibfield  {title} {\bibinfo {title} {General relativistic boson
  stars},\ }\href {https://doi.org/10.1088/0264-9381/20/20/201} {\bibfield
  {journal} {\bibinfo  {journal} {Class. Quantum Grav.}\ }\textbf {\bibinfo
  {volume} {20}},\ \bibinfo {pages} {R301} (\bibinfo {year}
  {2003})}\BibitemShut {NoStop}%
\bibitem [{\citenamefont {Kramer}\ \emph {et~al.}(1980)\citenamefont {Kramer},
  \citenamefont {Stephani}, \citenamefont {MacCallum},\ and\ \citenamefont
  {Hert}}]{Kramer-1980}%
  \BibitemOpen
  \bibfield  {author} {\bibinfo {author} {\bibfnamefont {D.}~\bibnamefont
  {Kramer}}, \bibinfo {author} {\bibfnamefont {H.}~\bibnamefont {Stephani}},
  \bibinfo {author} {\bibfnamefont {M.~A.}\ \bibnamefont {MacCallum}},\ and\
  \bibinfo {author} {\bibfnamefont {E.}~\bibnamefont {Hert}},\ }\href@noop {}
  {\emph {\bibinfo {title} {Exact {S}olutions of {E}instein’s {F}ield
  {E}quations}}}\ (\bibinfo  {publisher} {Deutsche Verlag der Wissenschaften,
  Berlin/Cambridge University Press},\ \bibinfo {address} {Cambridge,
  England},\ \bibinfo {year} {1980})\BibitemShut {NoStop}%
\bibitem [{\citenamefont {Delgaty}\ and\ \citenamefont
  {Lake}(1998)}]{Delgaty-1998}%
  \BibitemOpen
  \bibfield  {author} {\bibinfo {author} {\bibfnamefont {M.}~\bibnamefont
  {Delgaty}}\ and\ \bibinfo {author} {\bibfnamefont {K.}~\bibnamefont {Lake}},\
  }\bibfield  {title} {\bibinfo {title} {Physical acceptability of isolated,
  static, spherically symmetric, perfect fluid solutions of einstein's
  equations},\ }\href
  {https://doi.org/https://doi.org/10.1016/S0010-4655(98)00130-1} {\bibfield
  {journal} {\bibinfo  {journal} {Comput. Phys. Commun.}\ }\textbf {\bibinfo
  {volume} {115}},\ \bibinfo {pages} {395} (\bibinfo {year} {1998})},\ \bibinfo
  {note} {{C}omputer algebra in Physics Research}\BibitemShut {NoStop}%
\bibitem [{\citenamefont {Oppenheimer}\ and\ \citenamefont
  {Volkoff}(1939)}]{Oppenheimer-39}%
  \BibitemOpen
  \bibfield  {author} {\bibinfo {author} {\bibfnamefont {J.~R.}\ \bibnamefont
  {Oppenheimer}}\ and\ \bibinfo {author} {\bibfnamefont {G.~M.}\ \bibnamefont
  {Volkoff}},\ }\bibfield  {title} {\bibinfo {title} {On {M}assive {N}eutron
  {C}ores},\ }\href {https://doi.org/10.1103/PhysRev.55.374} {\bibfield
  {journal} {\bibinfo  {journal} {Phys. Rev.}\ }\textbf {\bibinfo {volume}
  {55}},\ \bibinfo {pages} {374} (\bibinfo {year} {1939})}\BibitemShut
  {NoStop}%
\bibitem [{\citenamefont {Raghoonundun}\ and\ \citenamefont
  {Hobill}(2015)}]{Raghoonundun-2015}%
  \BibitemOpen
  \bibfield  {author} {\bibinfo {author} {\bibfnamefont {A.~M.}\ \bibnamefont
  {Raghoonundun}}\ and\ \bibinfo {author} {\bibfnamefont {D.~W.}\ \bibnamefont
  {Hobill}},\ }\bibfield  {title} {\bibinfo {title} {Possible physical
  realizations of the {T}olman {VII} solution},\ }\href
  {https://doi.org/10.1103/PhysRevD.92.124005} {\bibfield  {journal} {\bibinfo
  {journal} {Phys. Rev. D}\ }\textbf {\bibinfo {volume} {92}},\ \bibinfo
  {pages} {124005} (\bibinfo {year} {2015})}\BibitemShut {NoStop}%
\bibitem [{\citenamefont {Negi}\ and\ \citenamefont
  {Durgapal}(1999)}]{Negi-1999}%
  \BibitemOpen
  \bibfield  {author} {\bibinfo {author} {\bibfnamefont {P.~S.}\ \bibnamefont
  {Negi}}\ and\ \bibinfo {author} {\bibfnamefont {M.~C.}\ \bibnamefont
  {Durgapal}},\ }\bibfield  {title} {\bibinfo {title} {Stable {U}ltracompact
  {O}bjects},\ }\href {https://doi.org/10.1023/A:1018807219245} {\bibfield
  {journal} {\bibinfo  {journal} {Gen. Relativ. Gravit.}\ }\textbf {\bibinfo
  {volume} {31}},\ \bibinfo {pages} {13} (\bibinfo {year} {1999})}\BibitemShut
  {NoStop}%
\bibitem [{\citenamefont {Negi}\ and\ \citenamefont
  {Durgapal}(2001)}]{Negi-2001}%
  \BibitemOpen
  \bibfield  {author} {\bibinfo {author} {\bibfnamefont {P.}~\bibnamefont
  {Negi}}\ and\ \bibinfo {author} {\bibfnamefont {M.}~\bibnamefont
  {Durgapal}},\ }\bibfield  {title} {\bibinfo {title} {Relativistic
  supermassive stars},\ }\href {https://doi.org/10.1023/A:1002707730439}
  {\bibfield  {journal} {\bibinfo  {journal} {Astrophys. Space Sci.}\ }\textbf
  {\bibinfo {volume} {275}},\ \bibinfo {pages} {185} (\bibinfo {year}
  {2001})}\BibitemShut {NoStop}%
\bibitem [{\citenamefont {Moustakidis}(2017)}]{Moustakidis-2017}%
  \BibitemOpen
  \bibfield  {author} {\bibinfo {author} {\bibfnamefont {C.}~\bibnamefont
  {Moustakidis}},\ }\bibfield  {title} {\bibinfo {title} {The stability of
  relativistic stars and the role of the adiabatic index},\ }\href
  {https://doi.org/10.1007/s10714-017-2232-9} {\bibfield  {journal} {\bibinfo
  {journal} {Gen. Relativ. Gravit.}\ }\textbf {\bibinfo {volume} {49}},\
  \bibinfo {pages} {68} (\bibinfo {year} {2017})}\BibitemShut {NoStop}%
\bibitem [{\citenamefont {Narain}\ \emph {et~al.}(2006)\citenamefont {Narain},
  \citenamefont {Schaffner-Bielich},\ and\ \citenamefont
  {Mishustin}}]{Narain-06}%
  \BibitemOpen
  \bibfield  {author} {\bibinfo {author} {\bibfnamefont {G.}~\bibnamefont
  {Narain}}, \bibinfo {author} {\bibfnamefont {J.}~\bibnamefont
  {Schaffner-Bielich}},\ and\ \bibinfo {author} {\bibfnamefont {I.~N.}\
  \bibnamefont {Mishustin}},\ }\bibfield  {title} {\bibinfo {title} {Compact
  stars made of fermionic dark matter},\ }\href
  {https://doi.org/10.1103/PhysRevD.74.063003} {\bibfield  {journal} {\bibinfo
  {journal} {Phys. Rev. D}\ }\textbf {\bibinfo {volume} {74}},\ \bibinfo
  {pages} {063003} (\bibinfo {year} {2006})}\BibitemShut {NoStop}%
\bibitem [{\citenamefont {Colpi}\ \emph {et~al.}(1986)\citenamefont {Colpi},
  \citenamefont {Shapiro},\ and\ \citenamefont {Wasserman}}]{Colpi-1986}%
  \BibitemOpen
  \bibfield  {author} {\bibinfo {author} {\bibfnamefont {M.}~\bibnamefont
  {Colpi}}, \bibinfo {author} {\bibfnamefont {S.~L.}\ \bibnamefont {Shapiro}},\
  and\ \bibinfo {author} {\bibfnamefont {I.}~\bibnamefont {Wasserman}},\
  }\bibfield  {title} {\bibinfo {title} {Boson {S}tars: {G}ravitational
  {E}quilibria of {S}elf-{I}nteracting {S}calar {F}ields},\ }\href
  {https://doi.org/10.1103/PhysRevLett.57.2485} {\bibfield  {journal} {\bibinfo
   {journal} {Phys. Rev. Lett.}\ }\textbf {\bibinfo {volume} {57}},\ \bibinfo
  {pages} {2485} (\bibinfo {year} {1986})}\BibitemShut {NoStop}%
\bibitem [{\citenamefont {Rafiei~Karkevandi}\ \emph {et~al.}(2022)\citenamefont
  {Rafiei~Karkevandi}, \citenamefont {Shakeri}, \citenamefont {Sagun},\ and\
  \citenamefont {Ivanytskyi}}]{PhysRevD.105.023001}%
  \BibitemOpen
  \bibfield  {author} {\bibinfo {author} {\bibfnamefont {D.}~\bibnamefont
  {Rafiei~Karkevandi}}, \bibinfo {author} {\bibfnamefont {S.}~\bibnamefont
  {Shakeri}}, \bibinfo {author} {\bibfnamefont {V.}~\bibnamefont {Sagun}},\
  and\ \bibinfo {author} {\bibfnamefont {O.}~\bibnamefont {Ivanytskyi}},\
  }\bibfield  {title} {\bibinfo {title} {Bosonic dark matter in neutron stars
  and its effect on gravitational wave signal},\ }\href
  {https://doi.org/10.1103/PhysRevD.105.023001} {\bibfield  {journal} {\bibinfo
   {journal} {Phys. Rev. D}\ }\textbf {\bibinfo {volume} {105}},\ \bibinfo
  {pages} {023001} (\bibinfo {year} {2022})}\BibitemShut {NoStop}%
\bibitem [{\citenamefont {Shakeri}\ and\ \citenamefont
  {Karkevandi}(2024)}]{PhysRevD.109.043029}%
  \BibitemOpen
  \bibfield  {author} {\bibinfo {author} {\bibfnamefont {S.}~\bibnamefont
  {Shakeri}}\ and\ \bibinfo {author} {\bibfnamefont {D.~R.}\ \bibnamefont
  {Karkevandi}},\ }\bibfield  {title} {\bibinfo {title} {Bosonic dark matter in
  light of the nicer precise mass-radius measurements},\ }\href
  {https://doi.org/10.1103/PhysRevD.109.043029} {\bibfield  {journal} {\bibinfo
   {journal} {Phys. Rev. D}\ }\textbf {\bibinfo {volume} {109}},\ \bibinfo
  {pages} {043029} (\bibinfo {year} {2024})}\BibitemShut {NoStop}%
\bibitem [{\citenamefont {Agnihotri}\ \emph {et~al.}(2009)\citenamefont
  {Agnihotri}, \citenamefont {Schaffner-Bielich},\ and\ \citenamefont
  {Mishustin}}]{Agnihotri-2009}%
  \BibitemOpen
  \bibfield  {author} {\bibinfo {author} {\bibfnamefont {P.}~\bibnamefont
  {Agnihotri}}, \bibinfo {author} {\bibfnamefont {J.}~\bibnamefont
  {Schaffner-Bielich}},\ and\ \bibinfo {author} {\bibfnamefont {I.~N.}\
  \bibnamefont {Mishustin}},\ }\bibfield  {title} {\bibinfo {title} {Boson
  stars with repulsive self-interactions},\ }\href
  {https://doi.org/10.1103/PhysRevD.79.084033} {\bibfield  {journal} {\bibinfo
  {journal} {Phys. Rev. D}\ }\textbf {\bibinfo {volume} {79}},\ \bibinfo
  {pages} {084033} (\bibinfo {year} {2009})}\BibitemShut {NoStop}%
\bibitem [{\citenamefont {Rutherford}\ \emph {et~al.}(2023)\citenamefont
  {Rutherford}, \citenamefont {Raaijmakers}, \citenamefont
  {Prescod-Weinstein},\ and\ \citenamefont {Watts}}]{Watts-2023}%
  \BibitemOpen
  \bibfield  {author} {\bibinfo {author} {\bibfnamefont {N.}~\bibnamefont
  {Rutherford}}, \bibinfo {author} {\bibfnamefont {G.}~\bibnamefont
  {Raaijmakers}}, \bibinfo {author} {\bibfnamefont {C.}~\bibnamefont
  {Prescod-Weinstein}},\ and\ \bibinfo {author} {\bibfnamefont
  {A.}~\bibnamefont {Watts}},\ }\bibfield  {title} {\bibinfo {title}
  {Constraining bosonic asymmetric dark matter with neutron star mass-radius
  measurements},\ }\href {https://doi.org/10.1103/PhysRevD.107.103051}
  {\bibfield  {journal} {\bibinfo  {journal} {Phys. Rev. D}\ }\textbf {\bibinfo
  {volume} {107}},\ \bibinfo {pages} {103051} (\bibinfo {year}
  {2023})}\BibitemShut {NoStop}%
\bibitem [{\citenamefont {Pitz}\ and\ \citenamefont
  {Schaffner-Bielich}(2023)}]{Pitz-2023}%
  \BibitemOpen
  \bibfield  {author} {\bibinfo {author} {\bibfnamefont {S.~L.}\ \bibnamefont
  {Pitz}}\ and\ \bibinfo {author} {\bibfnamefont {J.}~\bibnamefont
  {Schaffner-Bielich}},\ }\bibfield  {title} {\bibinfo {title} {Generating
  ultracompact boson stars with modified scalar potentials},\ }\href
  {https://doi.org/10.1103/PhysRevD.108.103043} {\bibfield  {journal} {\bibinfo
   {journal} {Phys. Rev. D}\ }\textbf {\bibinfo {volume} {108}},\ \bibinfo
  {pages} {103043} (\bibinfo {year} {2023})}\BibitemShut {NoStop}%
\bibitem [{\citenamefont {{Glendenning}}\ and\ \citenamefont
  {{Kettner}}(2000)}]{Glendenning-2000b}%
  \BibitemOpen
  \bibfield  {author} {\bibinfo {author} {\bibfnamefont {N.~K.}\ \bibnamefont
  {{Glendenning}}}\ and\ \bibinfo {author} {\bibfnamefont {C.}~\bibnamefont
  {{Kettner}}},\ }\bibfield  {title} {\bibinfo {title} {{Possible third family
  of compact stars more dense than neutron stars}},\ }\href
  {https://ui.adsabs.harvard.edu/abs/2000A%26A...353L...9G/abstract} {\bibfield
   {journal} {\bibinfo  {journal} {Astron. Astrophys.}\ }\textbf {\bibinfo
  {volume} {353}},\ \bibinfo {pages} {L9} (\bibinfo {year} {2000})}\BibitemShut
  {NoStop}%
\end{thebibliography}%

\end{document}